\documentclass[11pt]{report}
\usepackage[english]{babel}
\usepackage[latin1]{inputenc}
\usepackage{booktabs}
\usepackage{graphicx}
\usepackage{graphics}
\usepackage{caption}
\usepackage{amsmath}
\usepackage{amssymb}
\usepackage{amsthm}
\usepackage{array}
\usepackage{color}
 \usepackage[]{tabularx}
 
\usepackage[colorlinks,linkcolor=red!80!black,
citecolor=red!80!black,pdfpagelabels,hyperindex=true]{hyperref}
\usepackage{float}
\usepackage{tikz}
\usepackage{stmaryrd}
\usepackage{amsfonts}
\hypersetup{
    colorlinks=true,       
    linkcolor=brown!80!black,  
    citecolor=green!50!black,        
    filecolor=magenta,      
    urlcolor=red!80!black           
}

\theoremstyle{plain}

\theoremstyle{definition}

\theoremstyle{remark}

\title{Deep Learning  applied to Road Traffic Speed forecasting}

\author{ Thomas Epelbaum$^{a}$,Fabrice Gamboa$^{b}$, Jean-Michel Loubes$^{b}$ and Jessica Martin$^{c}$\footnote{Corresponding author: thomas.epelbaum@mediamobile.com}\\
\\
$^{a}$\textit{Mediamobile}, $^{b}$\textit{Institut de Math\'ematiques de Toulouse} and  $^{c}$\textit{INSA}}

\numberwithin{equation}{section}

\begin{document}
\def\layersep{2.5cm}

\maketitle

\begin{abstract}
In this paper, we propose  deep learning architectures (FNN, CNN and LSTM)  to forecast a regression model for time dependent data. These algorithm's are designed to handle  Floating Car Data (FCD) historic speeds to predict road traffic data. For this we aggregate the speeds into the network inputs in an innovative way. We compare the RMSE thus obtained with the results of a simpler physical model, and show that the latter achieves better RMSE accuracy. We also propose a new indicator, which evaluates the algorithms improvement when compared to a benchmark prediction. We conclude by questionning the interest of using deep learning methods for this specific regression task.

\vspace{0.3cm}

\noindent \textit{AMS subject classifications:} 

\vspace{0.3cm}

\end{abstract}

\tableofcontents

\chapter{ Road Traffic Speed forecasting: Modelization, Results}

\section{Introduction}

Traffic congestion is one of the major downsides of our ever-growing cities. The inconvenience for individuals stuck in traffic jams can sometimes be counted in hours per day, and weeks per year.  In this context, a road traffic speed forecasting algorithm could have a highly beneficial impact: it could feed a Dynamic Routing System (DRS), and allow one to anticipate the formation and the resorption of congestion. This could lead to intelligent recommandation to drivers, and point towards public measures for shifted departures to and from work (monetary incentives to companies/individuals...)


The speeds measured on the road network can be seen as a spatio-temporal time series. Although many models whether deterministic or stochastic have been created so far, these are either too time consuming to be used in real time, too demanding in terms of storage, or give too poor a result to prove valuable. Indeed the speeds present complex behaviors including seasonality, time dependency, spatial dependency and  drastic quick changes of patterns, making it uneasy to forecast. Yet since the evolution of the speeds rely on a real behavior, these time series present strong correlations and causality links that must be found.

In this paper, we focus on the design and implementation of Neural Networks (NN) to handle the forecast of time series applied to the case of road traffic. Using NN allows one to include a large number of explicative variables inside the model to capture the complexity of the time series. For this we study three kinds of NN architectures -- Feedforward (FNN), Convolutional (CNN) and Recurrent (RNN) -- to deal with the road traffic speed forecasting task.

There has been a growing interest for deep learning in the recent years, see for instance \cite{lecun2015deep} and references therein. Among all different kind of networks,  CNN are mostly used to classify images as in \cite{lecun1995convolutional} or \cite{zeiler2014visualizing}, and RNN are mostly used for Natural Language Processing tasks \cite{mikolov2010recurrent}, where one tries to guess the next word in a sentence given the previous ones. Yet for a regression task when dealing with time dependent data, few results exist in the statistical literature. 

In this work, we propose a way of feeding the NN to enhance the causality due to the seasonality of the observations which will be compared to the usual one. In addition we propose specific designs of the neural networks to forecast the data, each adapted to the type of NN used in this study.  Moreover, to evaluate the quality of the forecast, in addition to the standard Root Mean Square Error (RMSE), we propose a new $Q^2$ indicator, conceived so as to assess the quality of the model for time series forecast, completing the information conveyed by RMSE. Within this framework, we show that the Convolutional network severely underperforms the other NN variants and give some reasons why this may be. We also present a simple physical model, embedding the seasonality of the road traffic, that outperforms all the deep learning techniques while using order of magnitudes less parameters.

The paper divides as follow. We present the speed data at our disposal in Section \ref{sec:Explain_Problem}, and explain how we feed them to the NN in Section \ref{sec:Input}. The new criterion used to evaluate the quality of the different models is explained in Section \ref{sec:Model_evaluation}. The models themselves -FNN, CNN, RNN-LSTM -- are presented in Section \ref{sec:Deep_Learning}. We finally present the setup of our different simulations and our numerical results in Section \ref{sec:Results}.

\section{Data} \label{sec:Data}

\subsection{Speeds from FCD measures}\label{sec:Explain_Problem}

We deal here with speed forecasting on a road network. A road network is an oriented graph. On each oriented edge a speed is computed (via a technique called Floating Car Data, or FCD, see for instance in ~\cite{iet:/content/conferences/10.1049/cp_20000103}). A road network has to be described by several oriented edges as the speed limit, the topology, the presence of a traffic light... change for different zones of the network. On a edge $l$ of a road network, the speed at time $\tau$ is denoted $\tilde{v}_{l\tau}$. In our study, the speeds are available every 3 minutes, so that $\tilde{v}_{l\tau+1}$ comes 3 minutes after~$\tilde{v}_{l\tau}$.

In practice, we will normalize speeds. This procedure helps to identify similar speed patterns: formation of congestion or its resorption , without being affected by the absolute speed values. Otherwise, a classification algorithm would for instance fail to cluster together two edges which never experience any traffic jams if their speed limits are respectively 50 and 130 kph. 
To normalize the speeds, one needs to know the so-called free flow speed  $v^{_{\rm FFS}}_{l\tau}$ as defined in \cite{Fazio2014}. The free flow speed on Paris Ring Road is roughly $65kph$. We will thus from now on deal with the series of observations  
$$
v_{l\tau}=\frac{ \tilde{v}_{l\tau}}{v^{_{\rm FFS}}_{l\tau}}\;. $$

In this paper, we will study the external Paris ring road: 396 oriented edges in our network.
\begin{figure}[H]
\begin{center}
\begin{tikzpicture}
\node at (0,0) {\includegraphics[scale=0.22]{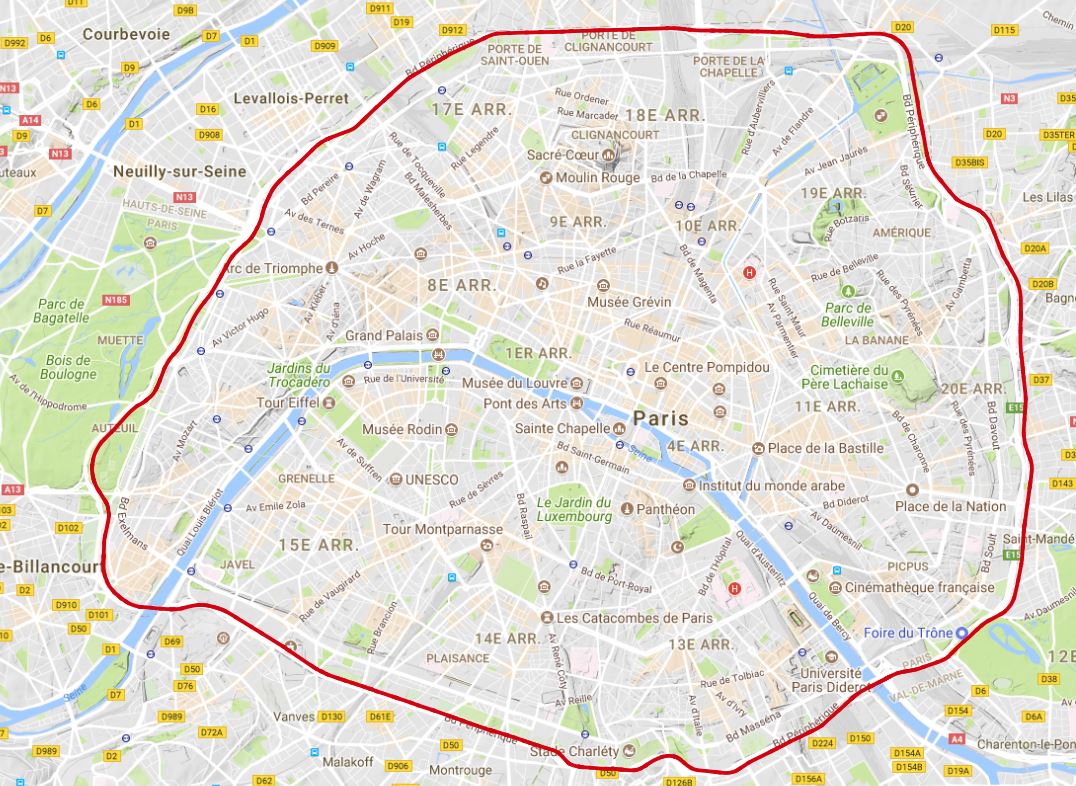}};
\end{tikzpicture}
\end{center}
\caption{\label{fig:zones} External Paris Ring Road.}
\end{figure}
 The training set will be data from the first 9 months of 2016, the test set belonging to tenth month of 2016. 
Note that to avoid predicting on trivial traffic states of constant speeds, we first removed the speed data from 11pm to 5 am of each day of our 10 month sample.

\subsection{Input speeds}\label{sec:Input}

In addition to the freedom of design in the FNN, CNN and RNN architectures, the way the past speeds are taken into account in the input of the networks play a crucial role in the results that one can obtain. Shall we take nearby edge past speed information, or can each edge predict its future speeds knowing only its past ones ? How far into the past shall we go ?  Hence the creation of the learning set determines the behaviour of the model and should be studied with care.

\vspace{0.2cm}

\vspace{0.2cm}

We will consider two kinds of input for the different NN architecture that we implement: a full input (large quantity of data) and a reduced one (supposed to capture the seasonality of the road traffic) 

\subsubsection{Full input} \label{sec:fullinput}

\begin{figure}[H]
\begin{center}
\begin{tikzpicture}
\node at (0,0) {\includegraphics[scale=0.2]{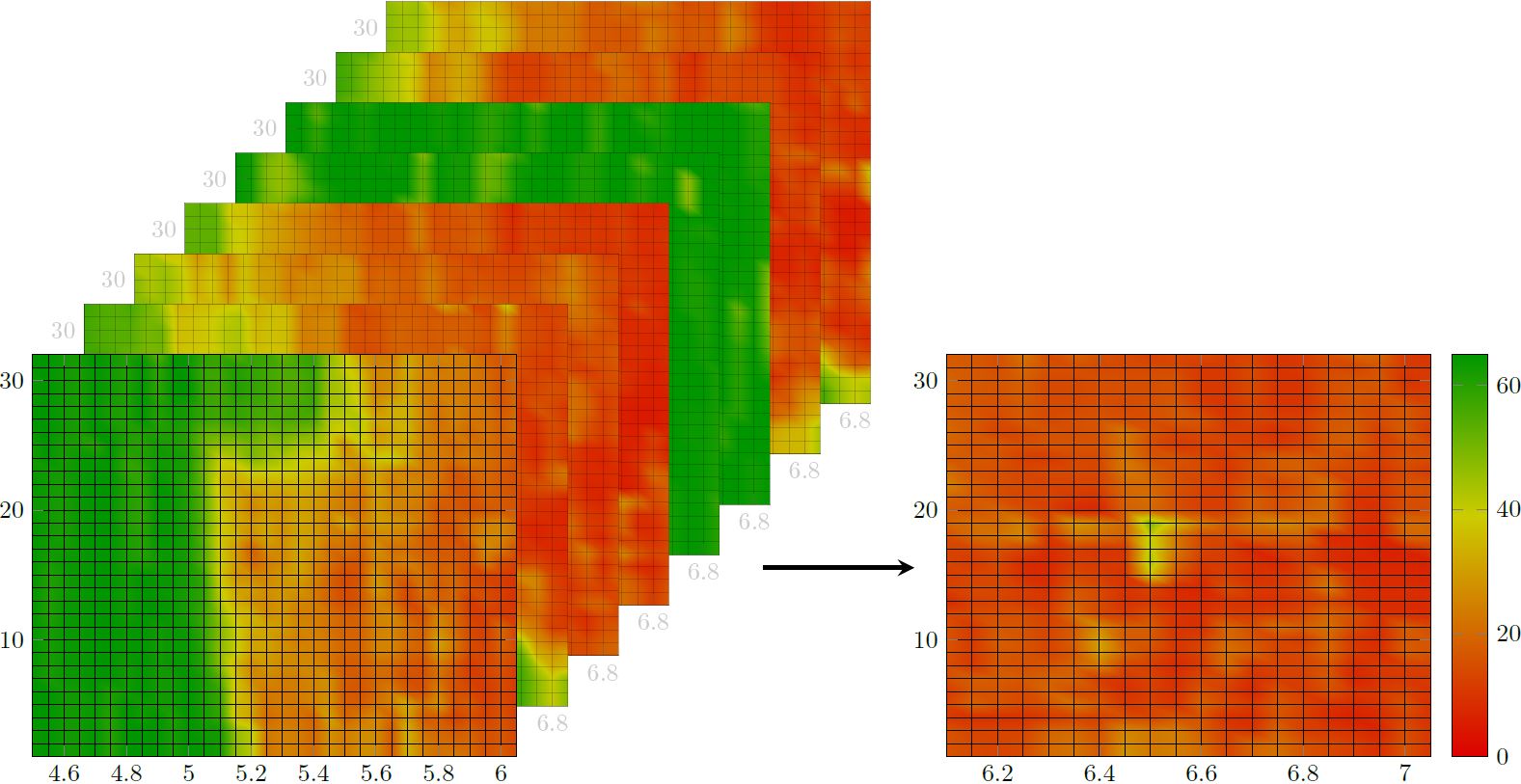} };
\end{tikzpicture}
\end{center}
\caption{\label{fig:Input32x32x8} full input used for the different NN architectures (left). Prediction example (right)}
\end{figure}

The full input of the NN corresponds to data on the current day (where one is trying to predict future speeds) and the immediate past $\mathcal{D}$ days. On these $\mathcal{D}+1$ days we consider $N_0$ contiguous edges of our graph. Starting at an arbitrary edge $N^{(t)}$ and an arbitrary time $T^{(t)}$, the full input of the network is then
\begin{align}
&\left(\left\{v_{_{N^{(t)}+l\,T^{(t)}-b}}\right\},\left\{v_{_{N^{(t)}+l\,T^{(t)}-\delta D+p}}\right\}\right)\;,
\label{eq:fullinp}
\end{align} 
with
\begin{align*}
l&\in \llbracket 0, N_0-1\rrbracket\;,&
b&\in \llbracket 1, \mathcal{B}_f\rrbracket\;,\notag\\
\delta&\in \llbracket 1,\mathcal{D}_f\rrbracket\;,&
p& \in  \llbracket -\mathcal{P}_{1f}, \mathcal{P}_{2f}\rrbracket\;
\end{align*}
where the $f$ subscript stands for full.
The $t$ index represents the $t$ sample of our training set, and D is a constant representing a full day in 3 minute intervals (480). We are thus considering the $3\mathcal{B}_f$ minutes of speed data on the current day (just before $T^{(t)}$), and a time window of $3\mathcal{P}_{1f}$ minutes before $T^{(t)}$ and $3\mathcal{P}_{2f}$ minutes after $T^{(t)}$ on past days. In practice we take
\begin{align*}
 N_0&=\mathcal{B}_f=32\;,&
\mathcal{D}_f&=7\;,\notag\\
\mathcal{P}_{1f}&=15\;,&
\mathcal{P}_{2f}&=16\;.
\end{align*}
This corresponds to $F_{0f}$ input variables with
\begin{align*}
F_{0f}&=N_0\times \left(\mathcal{B}_f+\left[\mathcal{P}_{1f}+\mathcal{P}_{2f}+1\right]\mathcal{D}_f\right)=8192\;.
\end{align*}
The prediction task corresponding to this full input is
\begin{align*}
&\left\{v_{_{N^{(t)}+l\,T^{(t)}+h}}\right\}\;,
\end{align*} 
with $h\in \llbracket 0,\mathcal{H}_f-1\rrbracket\;,$ and in practice we will take $
\mathcal{H}_f=20\;,$  hence an output of size $N_0\times\mathcal{H}_f=640\;.$
The full input and the corresponding output is illustrated in Figure \ref{fig:Input32x32x8} for a given $t$ sample of the training set.

Here, the output of the network (on the right part of Figure \ref{fig:Input32x32x8}) is the subsequent 20 three minute time intervals of the $N_0$ links of the current instance $t$ of the training set (time between 6:06 and 7:03 am). On the left part of Figure \ref{fig:Input32x32x8}, the eight colored matrices correspond to the traffic on the current day (matrix in the foreground, time between 4:30 and 6:03 am) and the traffic on the seven previous days (matrices in the background, time between 5:18 and 6:51 am).

\subsubsection{Reduced input}\label{sec:reducedinput}

\begin{figure}[H]
\begin{center}
\begin{tikzpicture}
\node at (0,0) {\includegraphics[scale=0.45]{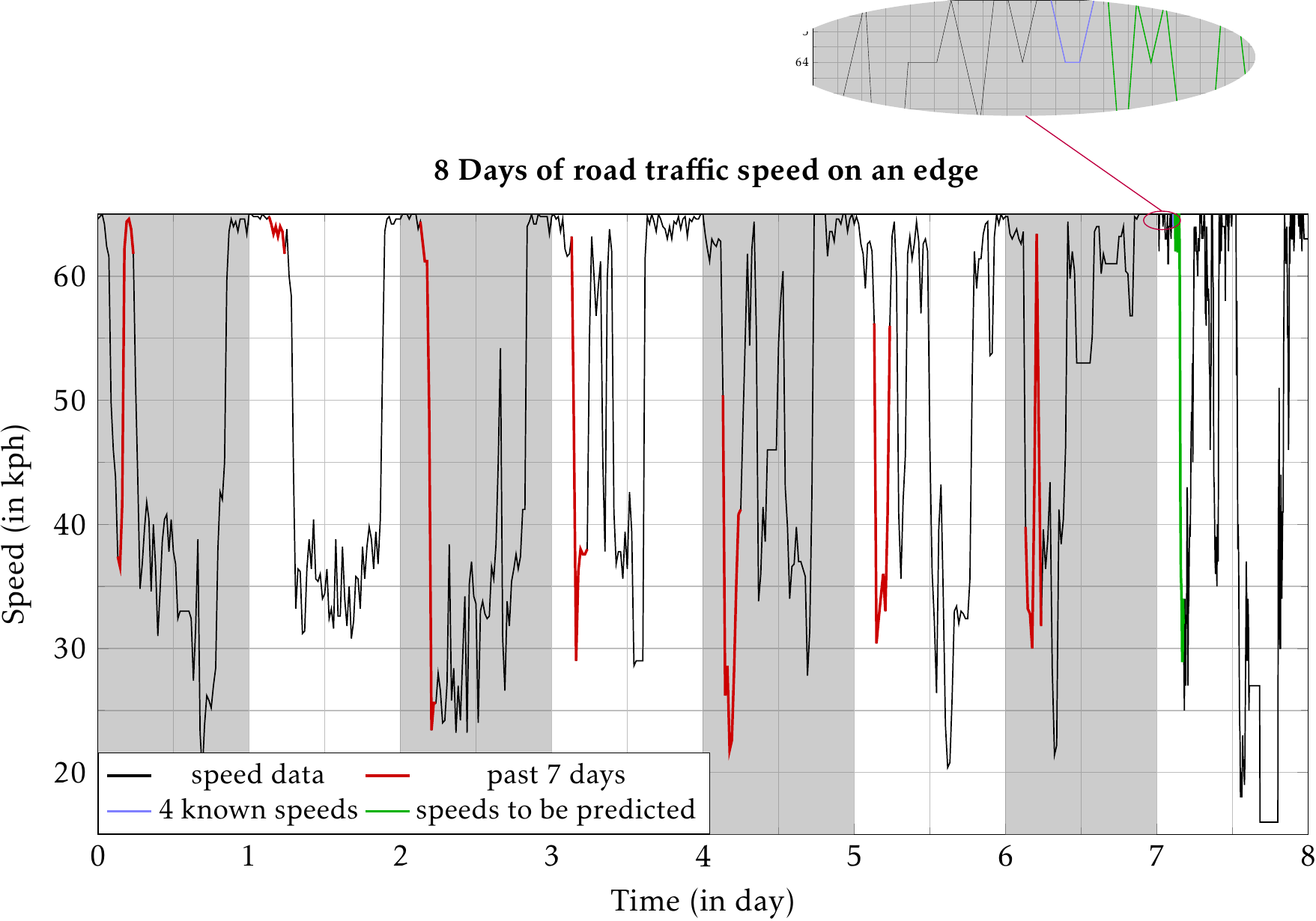}};
\end{tikzpicture}
\end{center}
\caption{\label{fig:for_seasonal_ar_synthesis} Reduced input example.}
\end{figure}

Starting back from our arbitrary edge $N^{(t)}$ and arbitrary time $T^{(t)}$, the reduced input deals with one edge. We thus forget the $l$ index in this part. The reduced input reads
\begin{align}
\left(\left\{v_{_{N^{(t)}\,T^{(t)}-b}}\right\},\left\{
\bar{v}_{_{N^{(t)}\,T^{(t)}-\delta D+p}}\right\}\right)\;. \label{eq:redinp}
\end{align} 
where (the $r$ subscript stands for reduced)
\begin{align*}
b&\in \llbracket 1, \mathcal{B}_r\rrbracket\;,&
\delta&\in \llbracket 1,\mathcal{D}_r\rrbracket\;,&
p& \in  \llbracket -\mathcal{P}_{1r}, \mathcal{P}_{2r}\rrbracket\;.
\end{align*}
The signification of $b$ and $\delta$ are unchanged (see Section \ref{sec:fullinput}). In practice we will take
 \begin{align*}
\mathcal{B}_r&=4\;,&
\mathcal{D}_r&=7\;.
\end{align*}
However, the way in which the previous days are taken into account is changed as follows. Instead of taking high resolution but noisy 3 minutes intervals, we will average over $\mathcal{M}$ such intervals, so that
\begin{align*}
\bar{v}_{_{N^{(t)}\,T^{(t)}-\delta D+p}}&=\frac{1}{\mathcal{M}}
\sum_{i=0}^{\mathcal{M}-1}v_{_{N^{(t)}\,T^{(t)}-\delta D+\mathcal{M}p+i}}\;.
\end{align*}
In practice we will take
\begin{align*}
\mathcal{M}&=5\;,&
\mathcal{P}_{1r}&=0\;,&
\mathcal{P}_{2r}&=3\;.
\end{align*}
The reduced input will thus be of size
\begin{align*}
F_{0r}&=\mathcal{B}_r+\left[\mathcal{P}_{1r}+\mathcal{P}_{2r}+1\right]\mathcal{D}_f=32
\end{align*}
The corresponding prediction task will now be
$
\left\{v_{_{N^{(t)}\,T^{(t)}+h}}\right\}\;,$ with  $h\in \llbracket 0,\mathcal{H}_r-1\rrbracket\;,$
and in practice we will take
$\mathcal{H}_r=20\;,$ and the output will be of size $\mathcal{H}_r=20$. The reduced input and its corresponding output is illustrated in Figure \ref{fig:for_seasonal_ar_synthesis}

On Figure \ref{fig:for_seasonal_ar_synthesis}, the traffic is represented for 8 contiguous days on the  $N^{(t)}$'th link of the Paris road network. The first 7 days (x axis between 0 and 7) are considered to be past days, and are pictured averaged per quarter-hours (corresponding to our $\mathcal{M}$ choice). The 8th day (x axis between 7 and 8) is depicted by 3 minute intervals.  $T^{(t)}$ characterizes the starting time at which one tries to predict afterwards. In Figure \ref{fig:for_seasonal_ar_synthesis}, this corresponds to the green part of the curve (see main plot and inset). The $v_{_{N^{(t)}\,T^{(t)}-b}}$ are the blue part of the curve (see both main plot and inset) while the $\bar{v}_{_{N^{(t)}\,T^{(t)}-\delta D+p}}$ are shown in red. 

\section{Model evaluation} \label{sec:Model_evaluation}

%

\subsection{RMSE and Q-score} \label{sec:RMSE_vs_Q-score}

Generally, in the speed forecasting paradigm\cite{Deng:2016:LSM:2939672.2939860},\cite{Fouladgar2017ScalableDT},\cite{MiwaTYM2004},\cite{SunHongyu}\cite{MaDaiHe:2017},\cite{Yuanchang2010} one evaluates the quality of a prediction algorithm considering the Root Mean Square Error (RMSE)
\begin{align*}
{\rm RMSE}&= \sqrt{\frac{1}{LT} \sum_{h=0}^{T}\sum_{l=1}^{L}\left(v_{l\tau+h}-\hat{v}_{l\tau+h}\right)^2}
\end{align*}
where $\hat{v}_{l\tau}$ is the speed predicted by the algorithm under consideration, $v_{l\tau+h}$ the ground truth speed, $T$ (19 in our study) covering all the time intervals where the prediction is taking place, and $L$ all the links of the road network considered. In our opinion, this indicator is not sharp enough to correctly assess the quality of a prediction algorithm. Indeed, for a constant time series, any RMSE close to 0 could fool people into thinking that the prediction algorithm is a good one, despite being worse than the prediction of taking a constant. At the other extreme, for a widely and swiftly changing time series, a large RMSE does not necessarily imply that the algorithm is a poor one.

\vspace{0.2cm}

To paliate this apparent paradox, it is necessary to introduce a new benchmark prediction. In our study we chose the real time propagation benchmark RTPB (taking $v_{l\tau}$ for any future speed $v_{l\tau+h}$), which is what is widely used in industrial DRS described for instance in \cite{fleischmann2004dynamic}. An algorithm with a low RMSE is not worth much interest if it constantly predicts worse than this simple benchmark, and an algorithm with a large RMSE but constantly beating the benchmark may be worth considering; it might just be that the road network under investigation is experiencing a large change of speed. We thus introduce
\begin{align}
Q^2&= 1-\frac{{\rm RMSE}^2}{{\rm RMSE}_{_{\rm bench}}^2}\;.\label{eq:Qbench}
%
%
\end{align}
where
\begin{align*}
{\rm RMSE}_{_{\rm bench}}&=\sqrt{\frac{1}{LT} \sum_{h=0}^{T}\sum_{l=1}^{L}\left(v_{l\tau+h}-v_{l\tau}\right)^2}\;.
\end{align*}
This Q-score (or $Q^2$ in the following) quantifies the improvement (or deterioration) of the considered algorithm when compared with the RTPB prediction. If $Q^2>0$ then there is an improvement (a $Q^2$ equal to 1 would mean a perfect prediction), otherwise if $Q^2<0$ there is a deterioration. Without this $Q^2$ specification, any speed prediction on an almost constant speed road network could lead to a very low RMSE without nevertheless being good, and one might be tempted to choose these kind of networks to beat "state of the art" RMSE. We hasten to add that this is not what was done by the papers that we reviewed \cite{MaDaiHe:2017,Fouladgar2017ScalableDT}, but it is notwithstanding hard to evaluate a traffic prediction algorithm having in mind real life applications if one solely consider the RMSE. Some studies\cite{Deng:2016:LSM:2939672.2939860,6482260} use an alternative indicator: the Mean Absolute Percentage Error (MAPE). 
\begin{align*}
{\rm MAPE}&=\frac{100}{LT} \sum_{h=0}^{T}\sum_{l=1}^{L}\left|1-\frac{\hat{v}_{l\tau+h}}{v_{l\tau+h}}\right|\;.
\end{align*}
Not using any benchmark algorithm, the MAPE is however subject to the same drawbacks as the RMSE when used alone.

\subsubsection{Loss function} \label{sec:FNNloss}

The loss function is a critical component of a NN architecture. For a classification task, the cross-entropy loss function is a standard choice. But in this study, as we are trying to make quantitative speed predictions that could feed a DRS, we will take a quadratic loss function
\begin{align*}
J_t(\Theta)&=\sum_{l=0}^{N_0-1}\sum_{h=0}^{\mathcal{H}-1}
\left(\hat{v}_{_{N^{(t)}+l\,T^{(t)}+h}}-v_{_{N^{(t)}+l\,T^{(t)}+h}}\right)^2\;,
\end{align*}
where $v$ is the ground truth speed and $\hat{v}$ the output of the NN under consideration. See Section \ref{sec:Input} for more details on the other notations.

Note that considering this particular loss function  means that we are considering the speed forecasting problem as a regression task. 
\vspace{0.2cm}

$J_t$ corresponds to the loss error on one sample of the training set, and we still have to decide how to train the NN. We make the standard choice of mini-batch Stochastic Gradient Descent (SGD)  as in \cite{johnson2013accelerating}, at each epoch $\mathcal{E}$ --one iteration of the training procedure, we pick  $T_{{\rm mb}}$ samples of the training set, and compute the mini-batch loss 
\begin{align}
J(\Theta)&=\frac{1}{2T_{{\rm mb}}}\sum_{t=0}^{T_{{\rm mb}}-1}J_t(\Theta)\;.\label{eq:nakedloss}
\end{align}

To achieve better accuracy, we had to regularize the Loss function. This is achieved using an $\ell^1+\ell^2$ penalty, also known as elastic net penalty, see for instance in \cite{zou2005regularization}, detailed in Section \ref{sec:Elasticnet}.

In this study we picked $T_{{\rm mb}}=50$ and $\mathcal{E}\sim 5.10^4$. The value of $\mathcal{E}$ is indicative, as we performed early stopping in order to improve accuracy as it is often done by users of NN as in \cite{caruana2001overfitting} to overcome the issue of the convergence of the algorithm.

\section{The models} \label{sec:Deep_Learning}

We present the different NN studied, starting with their common blocks and ending with their specificities

\subsection{Common properties of the Networks}

\subsubsection{Output function}

As we deal with normalized speed, we constrain the output of our model $\hat{v}$ to be between zero and one. We hence picked an output function $o$ such that
\begin{align*}
o(x)&= \begin{cases}
0 &\text{if $x<0$}\\
x &\text{if $0\leq x\leq 1$}\\
1 & \text{if $x> 1$}
\end{cases}
\end{align*}

\subsubsection{Batch Normalization} \label{sec:BN}

Batch normalization (BN) is the most popular regularization procedure and consists in jointly normalizing the mini-batch sets per data types at each input of a NN layer, except for the input of the network itself. This is because we want to keep track of what the data represents, hence keep their mean and standard deviation untouched. It should be mentioned from the outset that BN is an empirical procedure, though it has been shown to drastically improve state of the art performance on classification tasks for instance on CIFAR/MNIST on many challenges. 

In the original paper\cite{Ioffe2015}, the authors argued that this step should be done after the weight averaging/convolution (WA/C) operation and before the non-linear activation (NAC).

However, this choice stands on no theoretical grounds, and defeats the purpose of presenting a standardardized input to each NN layer. In addition, the back-propagation rules are a bit more cumbersome to write with the WA/C-BN-NAC convention. We therefore opted for a WA/C-NAC-BN architecture for all the layers -- except of course the pooling ones where there is nor an NAC neither a BN -- presented in this paper.

For the technical details on our BN implementation, see Section 1.9 of the supplementary material.

\subsubsection{Elastic net} \label{sec:Elasticnet}

In practice, using equation (\ref{eq:nakedloss}) for the loss function leads to poor results: the different NN that we considered never reach a positive $Q^2$. We thus had to regularize the loss function. This we did by using both $\ell_1$ and $\ell_2$ penalties, a procedure which turns our NN to so-called elastic nets. Calling $\Theta^{(\nu)f}_{f'}$ the weight matrix between the $\nu$'th and the $\nu+1$'th layer of our NN, this procedure amounts to add the following terms to the loss function 
\begin{align*}
J_{{\rm reg}}(\Theta)&=J(\Theta)
+\frac{\lambda_{_{\ell_2}}}{2} \sum_{\nu=0}^N\sum_{f'=0}^{F_\nu-1}
\sum_{f=0}^{F_{\nu+1}-1}\left(\Theta^{(\nu)f}_{f'}\right)^2
+\lambda_{_{\ell_1}}\sum_{\nu=0}^N\sum_{f'=0}^{F_\nu-1}\sum_{f=0}^{F_{\nu+1}-1}\left|\Theta^{(\nu)f}_{f'}\right|\;.
\end{align*}
With $F_{\nu}$ ($F_{\nu+1}$) the size of the $\nu$'th ($\nu+1$'th) layer of the NN. These new terms play a role in the update rules of the weight matrices as shown in~\cite{Epelbaum2017}. 

In our numerical simulations, we took $\lambda_{_{\ell_1}}$ and $\lambda_{_{\ell_2}}$ to be in the range $10^{-4}-10^{-1}$, the best value being picked by cross-validation.

\subsubsection{No Dropout}

In our study, it turned out that dropout -- with the probability $p$ of retaining a unit varying between $0.2-0.8$ for both the input and hidden layers -- only slows down convergence without reducing the RMSE. We therefore chose to discard dropout from our architectures.

\subsubsection{Adam Optimizer}

For the mini-batch SGD used in backpropagation\cite{LeCun:1998:EB:645754.668382}, we used the Adam optimizer proposed for instance in~\cite{Kingma2014} which keeps track of both the weight gradient $\Delta^{\Theta}_{e}$ ($e\in \llbracket 0,\mathcal{E}-1\rrbracket$) and its square via two epoch dependent vectors $m$ and $v$. For a given weight matrix $\Theta$, we thus have
\begin{align*}
m_{{\rm e}}&= \beta_1 m_{{\rm e-1}}+ (1-\beta_1)\Delta^{\Theta}_{e}\;,\notag\\
v_{{\rm e}}&= \beta_2 v_{{\rm e}}+ (1-\beta_2)\left(\Delta^{\Theta}_{e}\right)^2\;,
\end{align*}
with $\beta_1$ and $\beta_2$ parameters respectively set to $0.9$ and $0.999$, as advocated in the original paper\cite{Kingma2014}. We also observed that the final RMSE is poorly sensitive to the specific $\beta$ values. The different weight matrices $\Theta$ are then updated thanks to ($\epsilon=10^{-8}$) 
$$
\Theta_e=\Theta_{e-1}-\frac{\eta }{\sqrt{v_{{\rm e}}+\epsilon}}m_{{\rm e}}\;.$$
This is the optimization technique used for all our NN, along side a learning rate decay
$ \eta_e =e^{-\alpha_0}\eta_{e-1}\;,$ where 
$\alpha_0$ determined by cross-validation, and $\eta_0$ is usually initialized in the range $10^{-3}-10^{-2}$. Without this weight decay, the NN perform poorly.

\subsubsection{Activation function}

We used Leaky-ReLU activation functions in all our simulations (slight improvement on the standard ReLU choice), see for instance in \cite{agostinelli2014learning}. We also tested without success the ELU choice.

\subsection{Specificities of each Network}

\subsubsection{Feedforward Neural Networks} \label{sec:FNN}

For our FNN, we considered both one and three hidden layer architectures, with the WA-NLA-BN structure advocated in Section \ref{sec:BN}. We added no bias to the WA equation, as the latter is handled by the BN procedure, see the supplementary material for more details. We took a unique size $F_{\rm H}$ for the hidden layers, and varied it in between $8-256$.

We did not consider ResNet architectures\cite{He2015} in this study. This might be a natural following step, but in the present state of the art techniques for regression task with NN, we are pessimistic on any improvement this might achieve. Indeed, as shown in  our results, our 1 hidden layer FNN outperforms its 3 layer cousin.

\paragraph{Input of the FNN}: To feed our FNN, we just stuck together either the full or the reduced input (see equations  (\ref{eq:fullinp}) and (\ref{eq:redinp})) into a matrix  $X^{(t)}_{f}$. Here $t$ is the mini-batch index and $f\in F_{0(fr)}$ for respectively the full and reduced inputs.

\paragraph{Weight initialization}: we used the standard prescription\cite{pmlr-v9-glorot10a}
$$
\left[\Theta^{(\nu)}\right]_{{\rm init}}=\sqrt{\frac{6}{F_\nu+F_{\nu+1}}}\times\mathcal{N}(0,1)\;.$$

\subsubsection{Convolutional Neural Networks} \label{sec:CNN}

In our CNN study, we implemented a 16 weights VGG architecture \cite{DBLP:journals/corr/SimonyanZ14a} (ResNet could be considered in a future study). The input feature map is of size $\mathcal{D}_f+1$, while the images composing the input are of size $\mathcal{B}_f\times N_0$ for the current day and $\left(\mathcal{P}_{1f}+\mathcal{P}_{2f}+1\right)\times N_0$ for the previous days. In practice this corresponds to eight $32\times 32$ images, and we therefore implemented the standard convolutions using $3\times 3$ receptive fields with $1\time 1$ strides and a padding equal to $1$ on each image edge. We use pooling of $2\times 2$ receptive fields with $1\time 1$ strides. The size of the hidden feature maps has been taken in between $32-512$, and the size of the subsequent fully connected layers in between  $256-512$. More details on the architecture can be found in Section 2 of the supplementary material.

\paragraph{Input of the CNN}: For the CNN, we only considered the full input introduced in Section \ref{sec:fullinput}. Figure \ref{fig:Input32x32x8} makes complete sense for CNNs, as its left part represents the different feature maps of the input. The latter is now a four dimensional tensor $X^{(t)}_{\delta l(pb)}$ with t the mini-batch index and
\begin{align*}
{\rm for}\;\delta&=0,\;X^{(t)}_{\delta lb}=v_{_{N^{(t)}+l\,T^{(t)}-b}}\;,\notag\\
{\rm for}\;\delta&\neq 0,\;X^{(t)}_{\delta lp}=v_{_{N^{(t)}+l\,T^{(t)}-\delta D+p}}\;,
\end{align*}

\paragraph{Weight initialization}: We used the same prescription as for FNN.

\subsection{Recurrent Neural Networks} \label{sec:RNN}

For our RNN, we used the LSTM variant \cite{xingjian2015convolutional} with no peepholes in the input, forget and output gates. The latter are taken to be standard logistic $\sigma$ functions, while the cell update as well as the hidden state update are taken to be $\tanh$ functions. We performed BN as in the Feedforward/Convolutional before each input of a hidden layer. We took one hidden layer in the spatial direction and $\mathcal{H}$ in the temporal one.

\paragraph{Input of the RNN-LSTM}: an additional subtlety arises for RNN: the input depends on the temporal direction $\tau$ of the network. For the full input, we just took ($f$ here stands for $(b,p,l,\delta)$ with $l=0$ for the reduced input)
$$
X^{(t)(\tau)}_{f}=\left(\left\{v_{_{N^{(t)}+l\,T^{(t)}+\tau-b}}\right\},\left\{v_{_{N^{(t)}+l\,T^{(t)}+\tau-\delta D+p}}\right\}\right)\;,$$
implying that the input of a current temporal layer has to be fed with the output of the previous ones. We did just so in practice.  We took the same $F_{\rm H}$ as for FNNs.

\paragraph{Weight initialization}: We used a diagonal prescription inspired from \cite{SocherEtAl2013:CVG} ($J$ is the unit matrix, $\varepsilon\sim 10^{-2}$)
$$
\left[\Theta^{(\nu)}\right]_{{\rm init}}=\frac12 \mathbb{I}+J\mathcal{N}(0,\varepsilon)\;.$$

\paragraph{No clipping}: Clipping \cite{gupta2015deep} did not improve our model performances. We thus removed it.

\section{Results}\label{sec:Results}

\begin{figure}[H]
\begin{center}
\begin{tikzpicture}
\node at (0,0) {\includegraphics[scale=0.99]{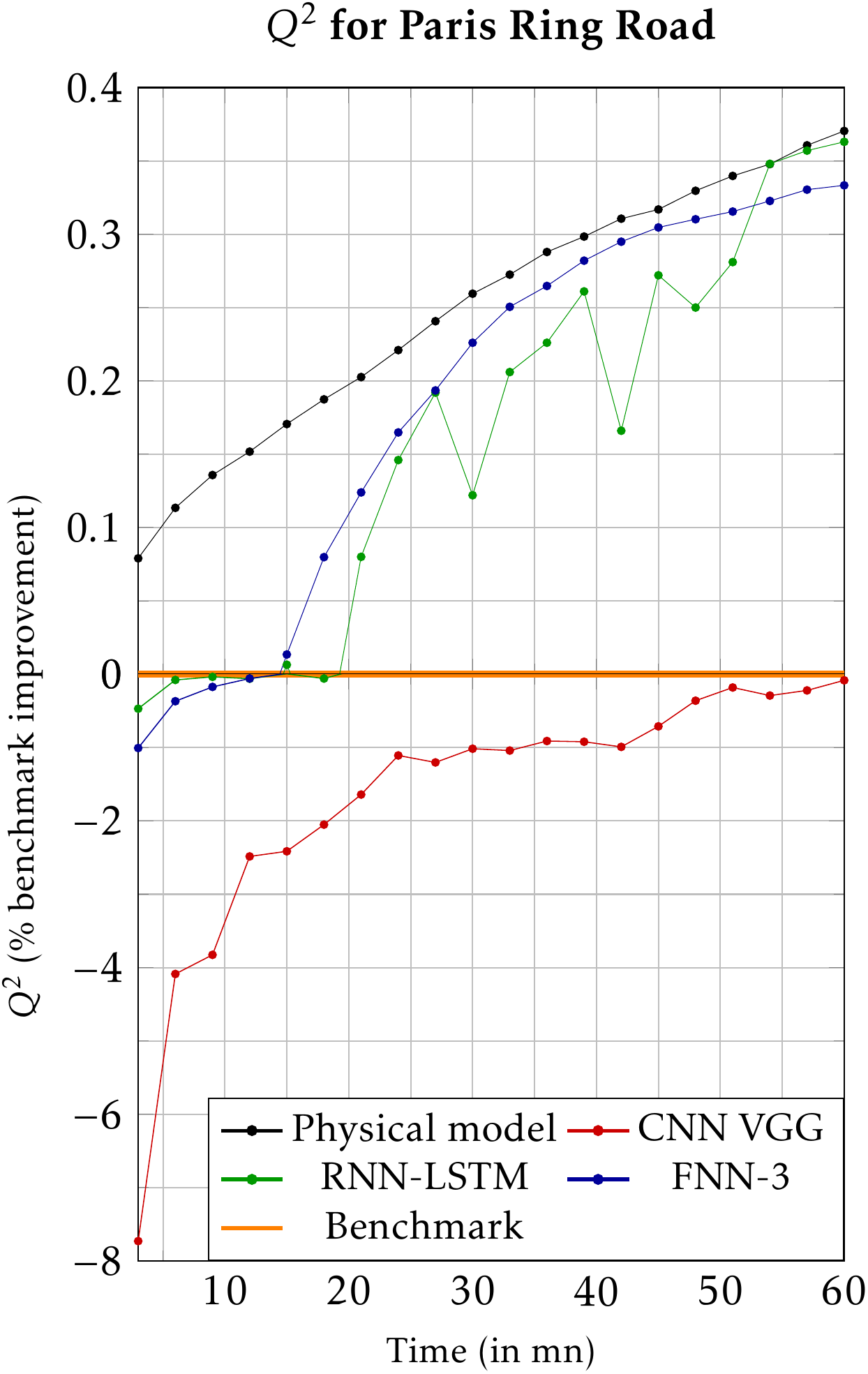}};
\end{tikzpicture}
\end{center}
\caption{\label{fig:DL_combined_Q2} Q-score for the different models studied.}
\end{figure}

With all the building blocks in place, we report here in Figures \ref{fig:DL_combined_RMSE} and \ref{fig:DL_combined_Q2} the compared performance of all the algorithms.

\subsection{Full input} \label{sec:Resfullinput}

\begin{figure}[H]
\begin{center}
\begin{tikzpicture}
\node at (0,0) {\includegraphics[scale=0.99]{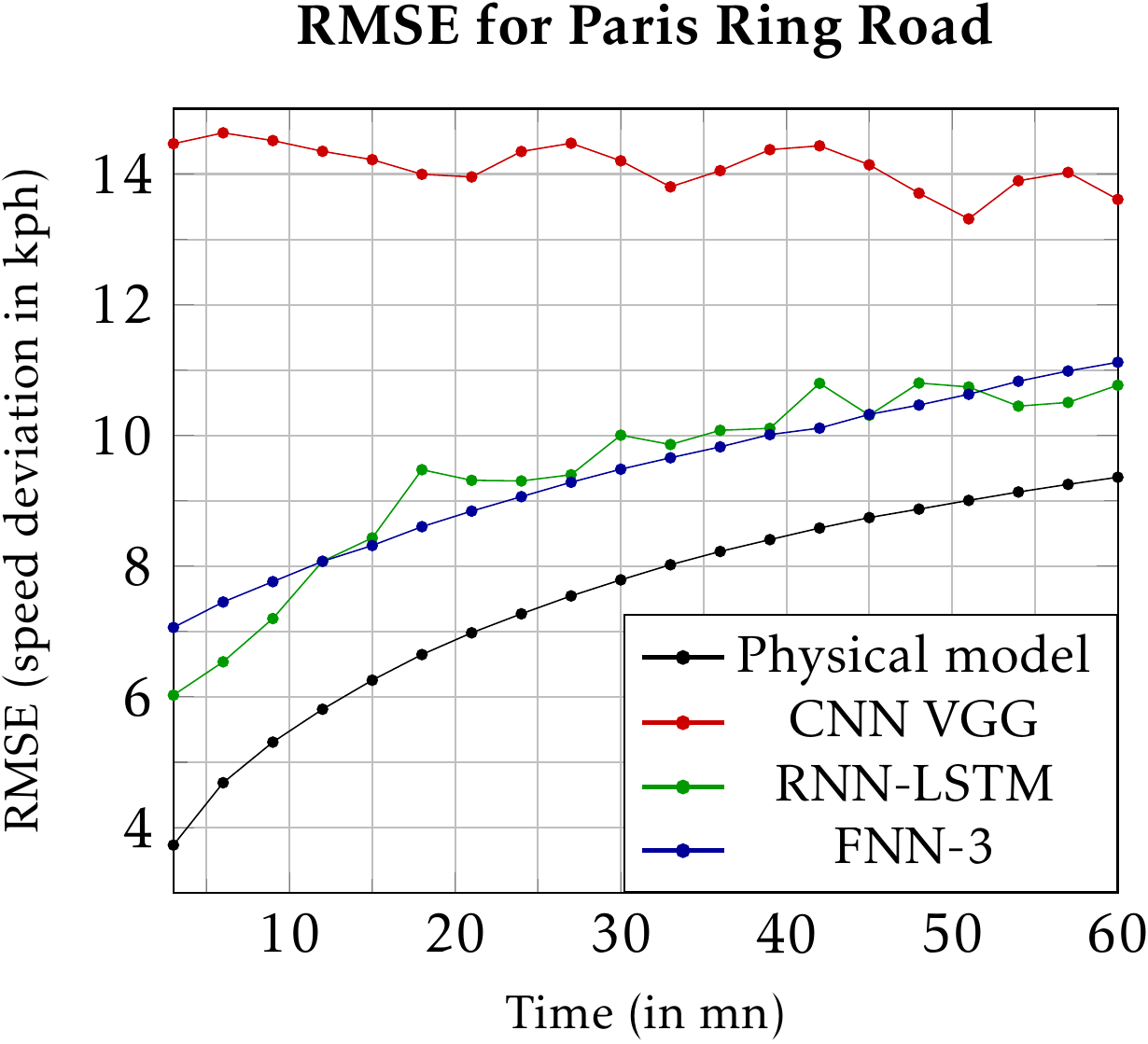}};
\end{tikzpicture}
\end{center}
\caption{\label{fig:DL_combined_RMSE} RMSE for the different models studied.}
\end{figure}

The RMSE of our different NN can be found for the full input in Figure \ref{fig:DL_combined_RMSE}. Here one can see several particular patterns.

\paragraph{Poor CNN performance}: the CNN severely underperforms the FNN and the RNN-LSTM. As the evolution of the car speeds is supposed to be a spatio-temporal field, considering the daily spatio-temporal velocities as images that characterize the traffic seemed to be a good frame to use CNN that proved helpful in pattern recognition. Yet road traffic speed prediction is not  a translationally invariant problem in the temporal direction and  no matter how the CNN weight matrices are tuned, the network assumes that the input image is translational invariant. The CNN hence fails to capture the fact that the right part of the input images in Figure \ref{fig:Input32x32x8} play a more important role that the left part, since it represents more recent traffic states.  
We extend our remark to any time series problem: the CNN seems ill equipped to handle prediction task properly in the case where there is strong causality in the time direction. Note however that another study in~\cite{MaDaiHe:2017} reports otherwise on some specific Chinese road networks, while only feeding the CNN with past traffic states of the current day (feature map of size 1 for the input layer). This may be due to some particular conditions in Chinese trafic. Yet the performance of NN are better than standard methods based on density models~\cite{CJS:CJS5550340307}.

\paragraph{Similar FNN/RNN-LSTM performance}: The two other kind of networks that we considered performed similarly, with a RMSE between $6-11kph$.

\paragraph{A simpler model performs better}: We developed a simpler physical model (PM) embedding the seasonality of the traffic. This model outperforms all NN, reaching state of the art RMSE value. Note that this work will be published in a future patent.

\paragraph{$Q^2$ utility}: We can assess the quality of our prediction when compared to the RTPB. We observe that FNN outperforms RNN-LSTM, despite having an almost identical RMSE. PM outperforms all NN, especially at early times.

\paragraph{Storage issues for real life applications}: the best FNN-3/VGG/LSTM models that we obtained have respectively $284672$/$243712$/$8224$ weight parameters, not counting the BN ones. Our best PM has $640$ parameters in total. In addition, it has a high degree of generalizability, allowing to mutualize the parameters for large road networks.

\subsection{Reduced input} \label{sec:Resreducedinput}
\begin{figure}[H]
\begin{center}
\begin{tikzpicture}
\node at (0,-11) {\includegraphics[scale=0.99]{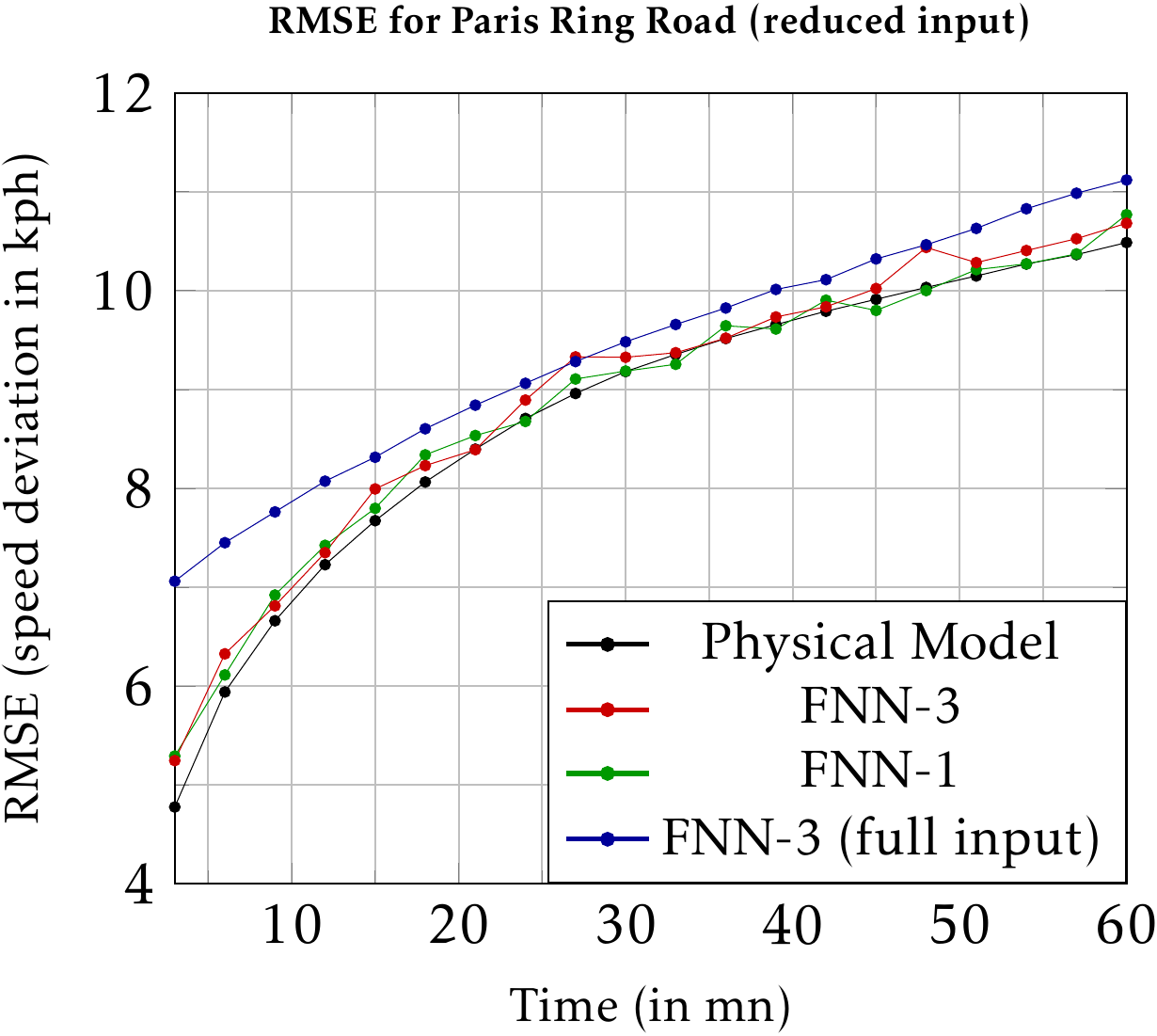}} ;
\end{tikzpicture}
\end{center}
\caption{\label{fig:Results_RMSE_Final}  RMSE for the reduced input}
\end{figure}

\begin{figure}[H]
\begin{center}
\begin{tikzpicture}
\node at (0,-11) {\includegraphics[scale=0.99]{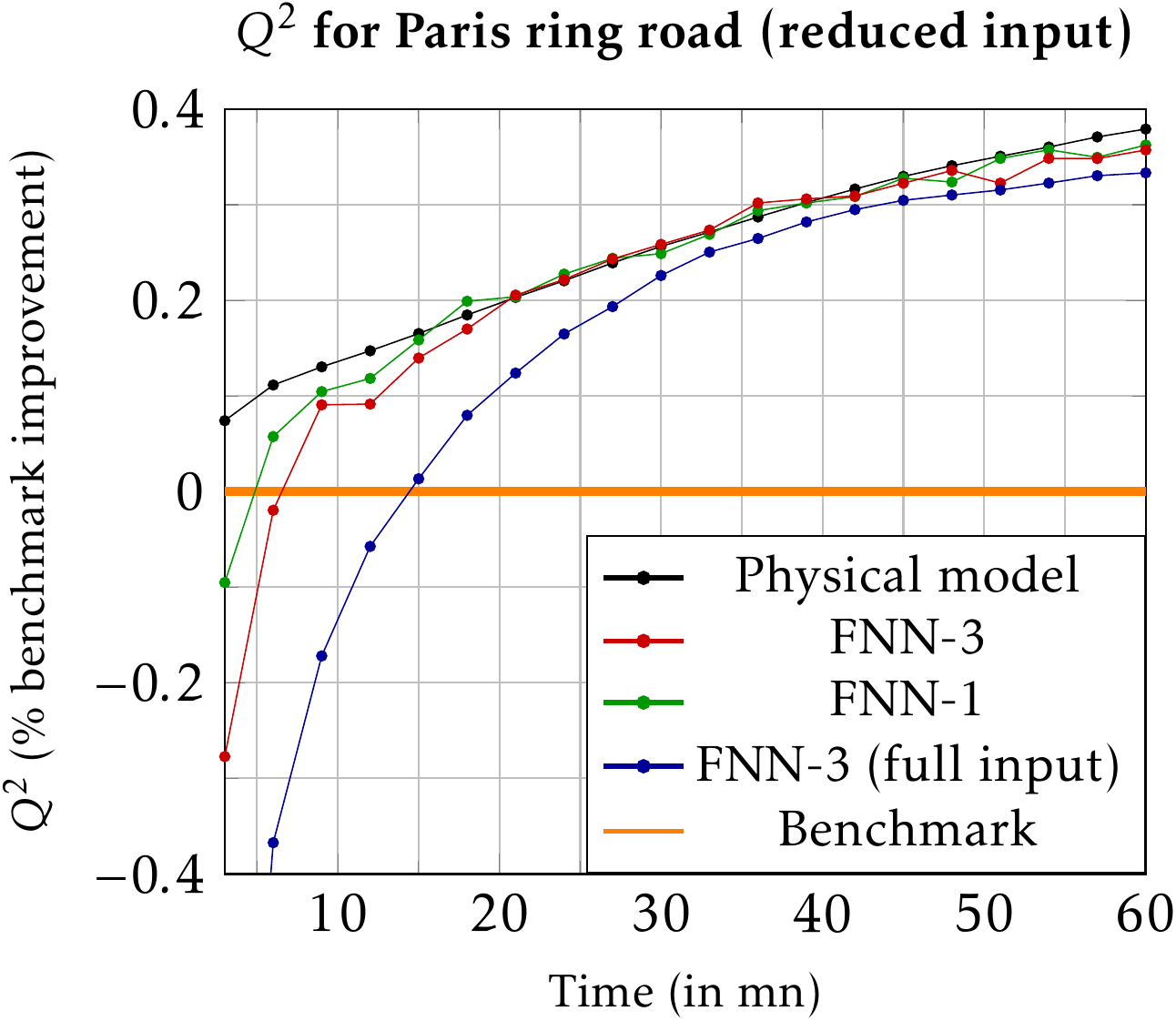}};
\end{tikzpicture}
\end{center}
\caption{\label{fig:Results_20_3_Final} Qbench for for the reduced inputs}
\end{figure}

We present in Figure \ref{fig:Results_RMSE_Final} and \ref{fig:Results_20_3_Final} the results obtained -- by two FNN architectures and the physical model -- for the reduced input. From these Figures we draw  
\paragraph{The reduced input outperforms the Full one}: The full input, though with a large number of explanatory variables, leads to worse results in terms both of $Q^2$ and RMSE. This might be due to either a poor initial condition choice -- we hope not -- or to the regression problem specificities.
\paragraph{The shallow FNN outperforms the deep one}: FNN-1 gives slightly better results than FNN-3.
\paragraph{The PM still outperforms NN:} but not by much.

\subsection{Importance of the Q-score} \label{sec:Q-score important}
\begin{figure}[H]
\begin{center}
\begin{tikzpicture}
\node at (0,0)  {\includegraphics[scale=0.99]{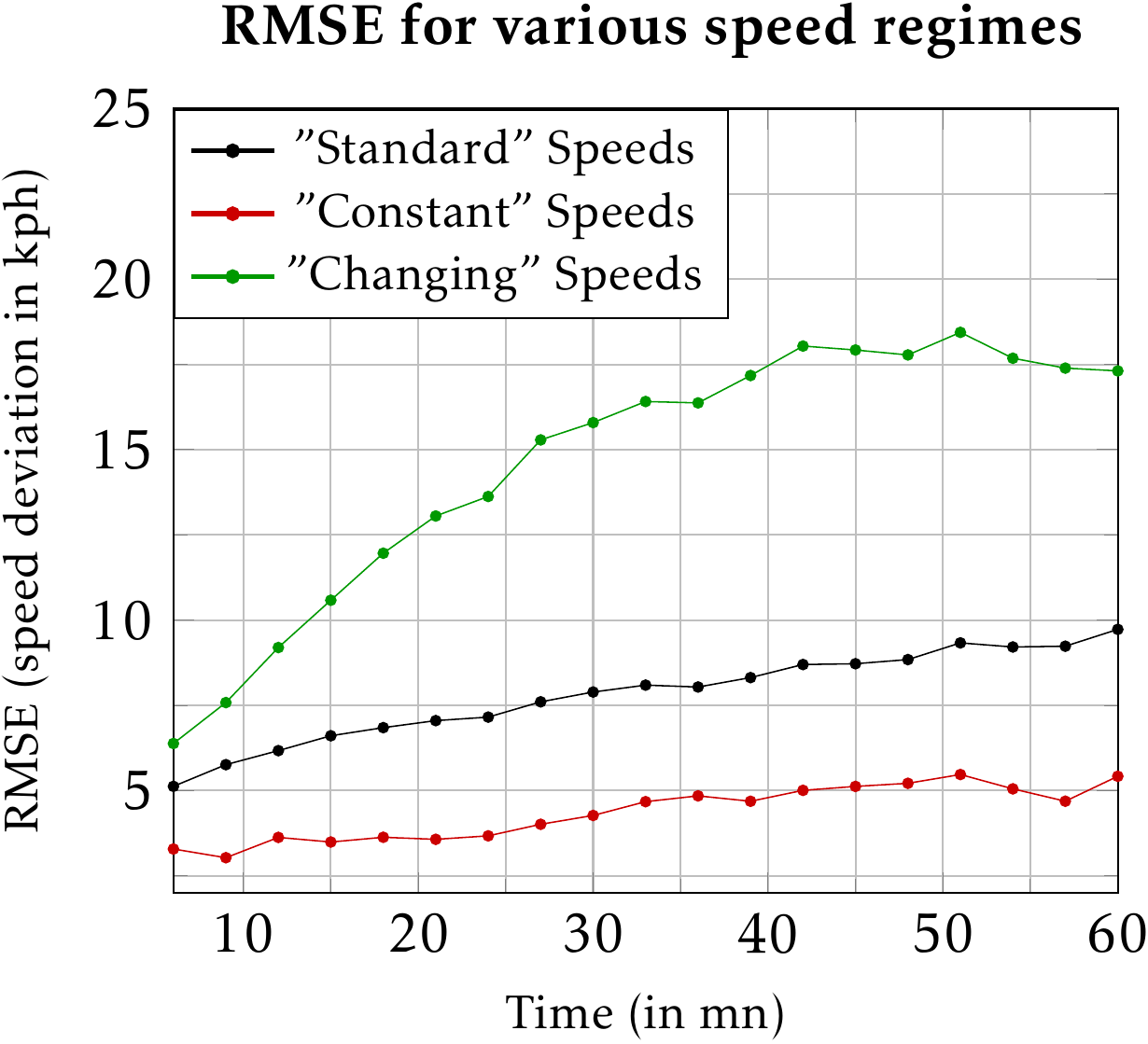}};
\end{tikzpicture}
\caption{\label{fig:Results_20_3_Various_regimes_RMSE}RMSE for different speed regimes}
\end{center}
\end{figure}

\begin{figure}[H]
\begin{center}
\begin{tikzpicture}
\node at (0,0)  {\includegraphics[scale=0.99]{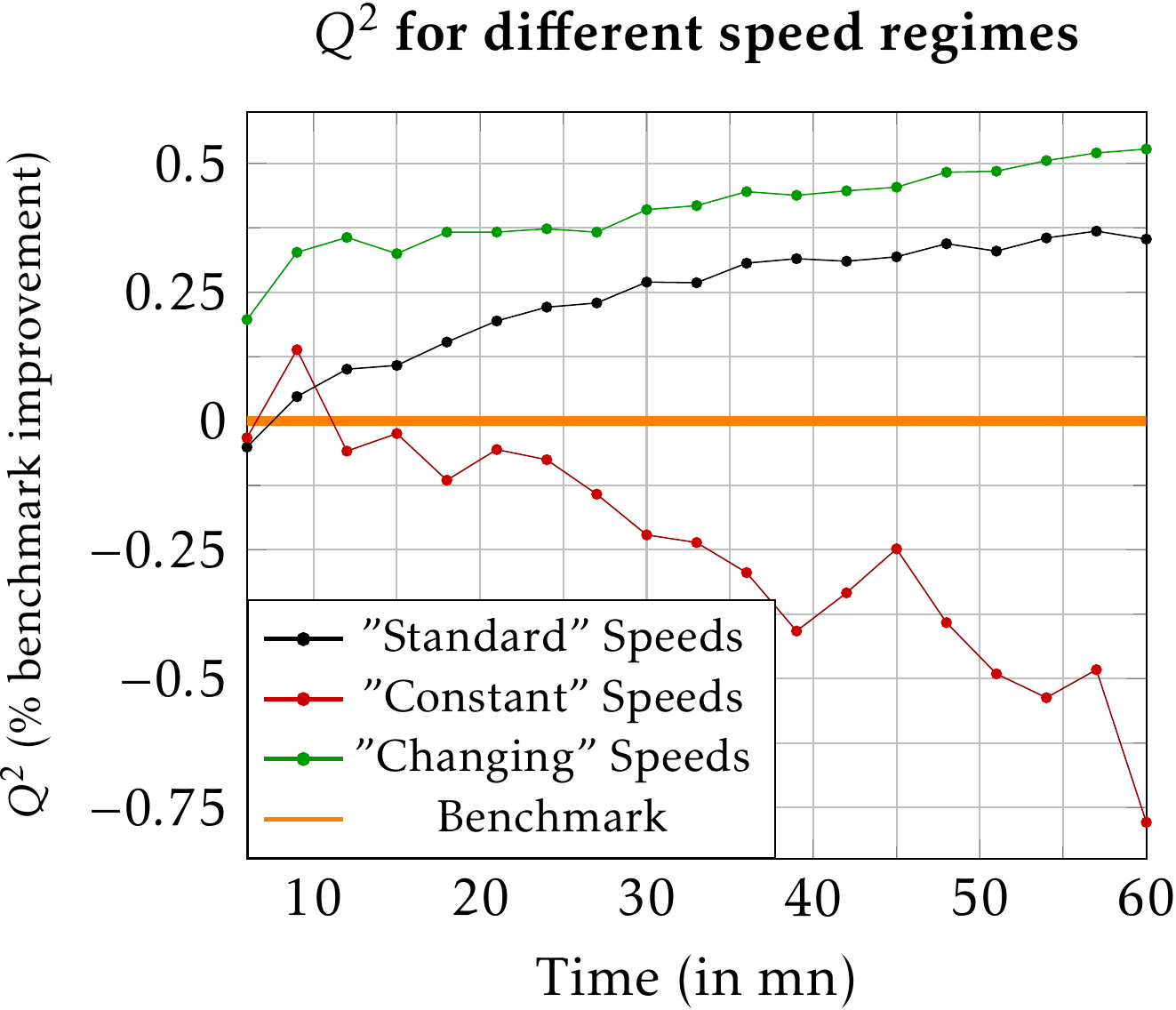}};
\end{tikzpicture}
\caption{\label{fig:Results_20_3_Various_regimes_Qbench}Q-score for different speed regimes}
\end{center}
\end{figure}

We have studied three speed regimes for the PM: the "constant" ("changing") one is the 10$\%$ part of the training set where the speeds before and after $T^{(t)}$ vary the less (more). The standard regime is the remaining 80$\%$. Results are shown in Figures \ref{fig:Results_20_3_Various_regimes_RMSE} and \ref{fig:Results_20_3_Various_regimes_Qbench}
\paragraph{Constant speeds lead to better RMSE}: The less the speed vary, the better the RMSE is, as expected.
\paragraph{Changing speeds lead to better $Q^2$}: 
Having low RMSE is not on its own a sign of a working model. The PM can't beat the Benchmark for "constant" speeds, while beating it by more than 50$\%$ after 50 minutes for the changing speeds. One should therefore jointly state RMSE and $Q^2$ in time series tasks.

\section{Conclusion} \label{sec:Conclusion}

We have implemented popular deep learning architectures, adapting their designs to the specific regression problem of predicting future road speeds, a generic example for time series presenting strong causality issues in both time and space. We showed that the CNN underperforms the other networks while  we built a PM that outperforms all NN architectures. We show that feeding the NN with more data leads to worst results, as does adding more layers to the FNN. We finally designed a new indicator, the $Q^2$, to be used jointly with the RMSE in time series problems.

\chapter{Deep Learning: supplementary and technical description}

\section{Feedforward Neural Networks} \label{sec:FNN}

Feedforward  networks (FNN) are an extension of the perceptron algorithm\cite{Rosenblatt58theperceptron:}. Their architecture is simple, but training them can be a daunting task. In the next few Sections, we  introduce the minimum FNN mathematical survival kit, stating what -- to the best of our knowledge -- are the empirical tricks and what stands on firmer mathematical ground.

\subsection{Some notations and definitions} \label{sec:FNNdef}

In the following, we call
\begin{itemize}
\item[$\bullet$] $N$ the number of layers (not counting the input) in the Neural Network.
\item[$\bullet$] $T_{{\rm train}}$ the number of training examples in the training set.
\item[$\bullet$] $T_{{\rm mb}}$ the number of training instances in a mini-batch (more on that later).  
\item[$\bullet$] $F_\nu$ the number of neurons in the $\nu$'th layer.
\item[$\bullet$] $t \in \llbracket0,T_{{\rm mb}}-1\rrbracket$ the mini-batch training instance index.
\item[$\bullet$] $\nu\in\llbracket0,N\rrbracket$ the index of the layer under consideration in the FNN. 
\item[$\bullet$] $h_{f}^{(t)(\nu)}$ where $f\in\llbracket0,F_\nu-1\rrbracket$ the neurons of the $\nu$'th layer. 
\item[$\bullet$] $X^{(t)}_f=h_{f}^{(t)(0)}$ where $f\in\llbracket0,F_0-1\rrbracket$ the input variables. 
\item[$\bullet$] $y^{(t)}_f$ where  $f\in[0,F_N-1]$ the output variables (to be predicted). 
\item[$\bullet$] $\hat{y}^{(t)}_f=h_{f}^{(t)(N)}$ where  $f\in[0,F_N-1]$ the output of the FNN. 
\item[$\bullet$] $\Theta_{f}^{(\nu)f'}$ for $f\in [0,F_{\nu}-1]$, $f'\in [0,F_{\nu+1}-1]$ and $\nu\in[0,N-1]$ the weight matrices.
\item[$\bullet$] A bias term can be included. In practice, we see when talking about the batch-normalization procedure that we can omit it, and we choose to do so in all our definitions.
\end{itemize}

\subsection{FNN architecture}

\begin{figure}[H]
\begin{center}
\begin{tikzpicture}
\node at (0,0) {\includegraphics[scale=1]{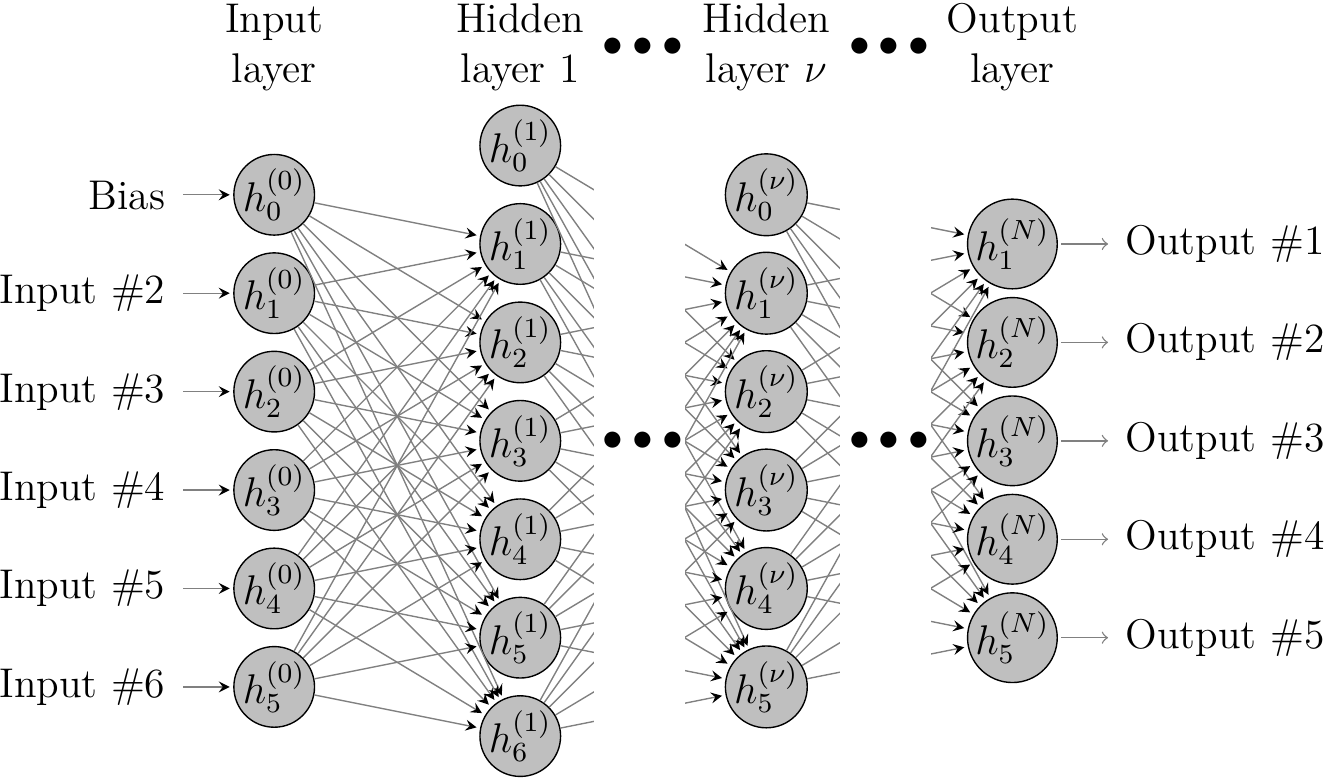}};
\end{tikzpicture}
\caption{\label{fig:1}Neural Network with $N+1$ layers ($N-1$ hidden layers). For simplicity of notations, the index referencing the training set has not been indicated. Shallow architectures (considered in the core of the paper) use only one hidden layer. Deep learning amounts to take several hidden layers, usually containing the same number of hidden neurons. This number should be in the ballpark of the average of the number of input and output variables.}
\end{center}
\end{figure}

A FNN is made of one input layer, one (shallow network) or more (deep network, hence the name deep learning) hidden layers and one output layer. Each layer of the network (except the output one) is connected to a following layer. This connectivity is central to the FNN structure and has two main features in its simplest form: a weight averaging feature and an activation feature. We review these features extensively in the following subsections

\subsection{Weight averaging}

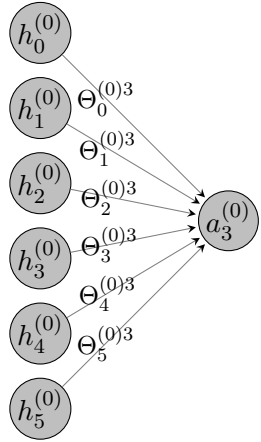
\begin{figure}[H]
\begin{center}
\begin{tikzpicture}[shorten >=1pt,-stealth,draw=black!50, node distance=\layersep,scale =1]
    \tikzstyle{every pin edge}=[stealth-,shorten <=1pt]
    \tikzstyle{neuron}=[circle,draw=black,fill=black!25,minimum size=17pt,inner sep=0pt]
    \tikzstyle{input neuron}=[neuron, fill=gray!50];
    \tikzstyle{output neuron}=[neuron, fill=gray!50];
    \tikzstyle{hidden neuron}=[neuron, fill=gray!50];
    \tikzstyle{annot} = [text width=4em, text centered]

    \foreach \name / \y in {1}
       	\pgfmathtruncatemacro{\m}{int(\y-1)}
        \node[input neuron] (I-\name) at (0,-\y) {$h_{\m}^{(0)}$};

    \foreach \name / \y in {2,...,6}
       	\pgfmathtruncatemacro{\m}{int(\y-1)}
        \node[input neuron] (I-\name) at (0,-\y) {$h_{\m}^{(0)}$};

    \foreach \name / \y in {4}
    	\pgfmathtruncatemacro{\m}{int(\y-1)}
        \path[yshift=0.5cm]
            node[hidden neuron] (H1-\name) at (\layersep,-\y cm) {$a_{\m}^{(0)}$};

 \path (I-1) edge node[pos=0.3,scale=0.9] {$\Theta^{(0)3}_0$} (H1-4);
 \path (I-2) edge node[pos=0.3,scale=0.9] {$\Theta^{(0)3}_1$} (H1-4);
 \path (I-3) edge node[pos=0.3,scale=0.9] {$\Theta^{(0)3}_2$} (H1-4);
 \path (I-4) edge node[pos=0.3,scale=0.9] {$\Theta^{(0)3}_3$} (H1-4);
 \path (I-5) edge node[pos=0.3,scale=0.9] {$\Theta^{(0)3}_4$} (H1-4);
 \path (I-6) edge node[pos=0.3,scale=0.9] {$\Theta^{(0)3}_5$} (H1-4);

\end{tikzpicture}
\caption{\label{fig:3}Weight averaging procedure.}
\end{center}
\end{figure}

One of the two main components of a FNN is a weight averaging procedure, which amounts to average the previous layer with some weight matrix to obtain the next layer. This is illustrated in Figure \ref{fig:3}

\vspace{0.2cm}

Formally, the weight averaging procedure reads:

\begin{align}
a_{f}^{(t)(\nu)}&=\sum^{F_\nu-1}_{f'=0}\Theta^{(\nu)f}_{\,f'}h^{(t)(\nu)}_{f'}
=\left(\Theta^{(\nu)T}h^{(t)(\nu)}\right)_f\;.
\end{align}

Here we have delibarately ommited a potential bias term, as it is handled in Section \ref{sec:FNNBN} where we talk about batch normalization. In practice, for all our numerical simulations, the weights are learned using the backpropagation procedure\cite{LeCun:1998:EB:645754.668382} with the Adam optimizer method for gradient descent\cite{Kingma2014}.

\subsection{Activation function} \label{sec:FNNReLU}

The hidden neuron of each layer is defined as
\begin{align}
h_{f}^{(t)(\nu+1)}&=g\left(a_{f}^{(t)(\nu)}\right)\;,
\end{align}
where $g$ is an activation function -- the second main ingredient of a FNN -- whose non-linearity allows the prediction of arbitrary output data. In practice, $g$ is usually taken to be either a sigmoid, a $\tanh$ a Rectified Linear Unit\cite{Hahnloser2000} (ReLU), or its variants: Leaky ReLU, ELU...\cite{Clevert2015FastAA}. The ReLU is defined as 

\begin{align}
g(x)&={\rm ReLU}(x)=\begin{cases} 
      x & x\geq 0 \\
      0& x<0 
   \end{cases}\;.
\end{align}
Its derivative is
\begin{align}
{\rm ReLU}'(x)&=\begin{cases} 
      1 & x\geq 0 \\
      0 & x<0 
   \end{cases}\;.
\end{align}

The choice of the activation function usually follows empirical tests, though it has been formally shown \cite{pmlr-v40-Choromanska15},\cite{Sagun2014} that Neural Networks cannot converge if the activation function is too complicated. With these building blocks in place, let us consider the different layers of the network one at a time.

\subsection{Input layer} \label{sec:FNNinput}

As explained in the core of the paper, we consider a full and a reduced input. But for all purposes here, we deal with an $h^{(t)(0)}_{f}=X^{(t)}_{f}$ input layer, with $f\in\llbracket0,F_0-1\rrbracket$ and $t \in \llbracket0,T_{{\rm mb}}-1\rrbracket$.

\subsection{Fully connected layer} \label{sec:FNNfc}

The fully connected operation is just the conjunction of the weight averaging and the activation procedure. Namely, for $\nu\in \llbracket 0,N-1 \rrbracket$
\begin{align}
a_{f}^{(t)(\nu)}&=\sum^{F_\nu-1}_{f'=0}\Theta^{(\nu)f}_{f'}h^{(t)(\nu)}_{f'}\;,
\end{align}
and for $\nu\in \llbracket 0,N-2 \rrbracket$
\begin{align}
h_{f}^{(t)(\nu+1)}&=g\left(a_{f}^{(t)(\nu)}\right)\;.
\end{align}
In the case where $\nu=N-1$, the activation function is replaced by an output function (see the next section).

\subsection{Output layer} \label{sec:FNNoutput}

The output of the FNN reads
 \begin{align}
h_{f}^{(t)(N)}&=o\left(a_{f}^{(t)(N-1)}\right)\;,
\end{align}
where $o$ is called the output function. In the case of the Euclidean loss function (se the next section), the output function is just the identity. Nevertheless, in our case we know that the normalized speeds cannot take values greater than 1 and smaller than 0, we therefore impose this constraint on the output of our FNN and take
\begin{align}
o(x)&=\begin{cases} 
      0 &x<0 \\
      x & 0\leq x\leq 1 \\
      1& x>1 
   \end{cases}\;.
\end{align}

\subsection{Loss function} \label{sec:FNNloss}

The loss function evaluates the error made by the FNN when it tries to estimate the data to be predicted. As explained in the core of the paper for a regression problem this is generally the mean square error (MSE)
\begin{align}
J(\Theta)&=\frac{1}{2T_{{\rm mb}}}\sum_{t=0}^{T_{{\rm mb}}-1}\sum_{f=0}^{F_N-1}
\left(y_f^{(t)}-h_{f}^{(t)(N)}\right)^2\;.
\end{align}
To obtain better results, one usually regularizes the Loss function. In addition to Batch normalization (explained in the next section), we also use $\ell_1$ and $\ell_2$ regularization. This amounts to add the following terms to the loss function 
\begin{align}
J_{{\rm reg}}(\Theta)&=J(\Theta)
+\frac{\lambda_{\ell_2}}{2} \sum_{\nu=0}^N\sum_{f'=0}^{F_\nu-1}\sum_{f=0}^{F_{\nu+1}-1}\left(\Theta^{(\nu)f}_{f'}\right)^2\notag\\
&+\lambda_{\ell_1}\sum_{\nu=0}^N\sum_{f'=0}^{F_\nu-1}\sum_{f=0}^{F_{\nu+1}-1}\left|\Theta^{(\nu)f}_{f'}\right|\;.
\end{align}
These new terms play a role in the update rules of the weight matrices\cite{Epelbaum2017}. 

\subsection{Batch Normalization} \label{sec:FNNBN}

As explained in the core of the paper, Batch normalization (BN) consists in jointly normalizing the mini-batch sets $T_{{\rm mb}}$ per data types $F_\nu$ at each input of a NN layer, except for the input of the network itself. In our case, we thus consider for $\nu \in \llbracket 0,N-2\rrbracket$
\begin{align}
\tilde{h}_{f}^{(t)(\nu)}&=\frac{h_{f}^{(t)(\nu+1)}-\hat{h}_{f}^{(\nu)}}
{\sqrt{\left(\hat{\sigma}_{f}^{(\nu)}\right)^2+\epsilon}}\;.
\end{align}
Here, $\hat{h}_{f}^{(\nu)}$ and $\hat{\sigma}_{f}^{(\nu)}$ are respectively the mean and the standard deviation of $h_{f}^{(t)(\nu+1)}$ with respect to the batch index $t$. To make sure that the Batch Normalization operation can also represent the identity transform, we standardly add two additional parameters $(\gamma_f,\beta_f)$ to the model (learned by backpropagation \cite{Epelbaum2017})
\begin{align}
y^{(t)(\nu)}_{f}&=\gamma^{(\nu)}_f\,\tilde{h}_{f}^{(t)(\nu)}+\beta^{(\nu)}_f\;.
%
\end{align}
The presence of the coefficient $\beta^{(\nu)}_f$ is what pushed us to get rid of the bias term. Indeed, it is now naturally included in batchnorm. During training, one must compute a running sum for the mean and the variance, that serve for the evaluation of the cross-validation and the test set. calling $e$ the current epoch
\begin{align}
\mathbb{E}\left[h_{f}^{(t)(\nu+1)}\right]_{e+1} &=
\frac{e\mathbb{E}\left[h_{f}^{(t)(\nu)}\right]_{e}+\hat{h}_{f}^{(\nu)}}{e+1}\;,\\
\mathbb{V}ar\left[h_{f}^{(t)(\nu+1)}\right]_{e+1} &=
\frac{e\mathbb{V}ar\left[h_{f}^{(t)(\nu)}\right]_{e}+\left(\hat{\sigma}_{f}^{(\nu)}\right)^2}{e+1}
\end{align}
and what is used at test time is
\begin{align}
\mathbb{E}\left[h_{f}^{(t)(\nu)}\right]&=\mathbb{E}\left[h_{f}^{(t)(\nu)}\right]\;,\notag\\
\mathbb{V}ar\left[h_{f}^{(t)(\nu)}\right]&=
\frac{T_{{\rm mb}}}{T_{{\rm mb}}-1}\mathbb{V}ar\left[h_{f}^{(t)(\nu)}\right]\;.
\end{align}
so that at test time 
\begin{align}
y^{(t)(\nu)}_{f}&=\gamma^{(\nu)}_f\,
\frac{h_{f}^{(t)(\nu)}-\mathbb{E}[h_{f}^{(t)(\nu)}]}{\sqrt{\mathbb{V}ar\left[h_{f}^{(t)(\nu)}\right]+\epsilon}}
+\beta^{(\nu)}_f\;.
\end{align}

\vspace{0.2cm}

In practice when using BN, and as advocated in the original paper\cite{Ioffe2015}, one can get rid without loss of precision of the former most popular regularization technique before BN introduction: Dropout\cite{Srivastava:2014:DSW:2627435.2670313}. We adopt this convention in the following, as our test experienced a loss of precision with the joint use of BN and dropout. This might be due to the peculiarity of our problem which is a regression and not a classification task. To the best of our knowledge, the litterature on NN for regression task is pretty scarce. Going back to the FNN structure, this induces the following change to the weight averaging formula of Section \ref{sec:FNNfc}: for $ \nu\in \llbracket 1,N-1\rrbracket$
\begin{align}
a_{f}^{(t)(\nu)}&=\sum^{F_\nu-1}_{f'=0}\Theta^{(\nu)f}_{f'}y^{(t)(\nu-1)}_{f'}\;.
\end{align}

\subsection{Architecture considered in practice}

Schematically denoting a hidden fully-connected unit as (with the WA/NLA/BN order advocated in the core of the paper)

\begin{figure}[H]
\begin{center}
\begin{tikzpicture}
\node at (0,0) {\includegraphics[scale=1.8]{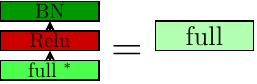}};
\end{tikzpicture}
\end{center}
\caption{\label{fig:fc_equiv} A FNN fully connected layer}
\end{figure}

and the FNN output unit as

\begin{figure}[H]
\begin{center}
\begin{tikzpicture}
\node at (0,0) {\includegraphics[scale=1.8]{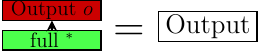}};
\end{tikzpicture}
\end{center}
\caption{\label{fig:output_equiv} The FNN output layer}
\end{figure}
\begin{figure}[H]
\begin{center}
\begin{tikzpicture}
\node at (0,0) {\includegraphics[scale=1.8]{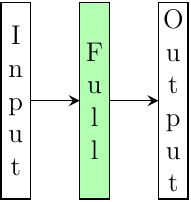}};
\end{tikzpicture}
\end{center}
\caption{\label{fig:FNN_network_1_layer} FNN with one hidden layer}
\end{figure}


we consider in the core of our paper the following two FNN architectures. A one hidden layer FNN depicted in Figure \ref{fig:FNN_network_1_layer}

\begin{figure}[H]
\begin{center}
\begin{tikzpicture}
\node at (0,0) {\includegraphics[scale=1.8]{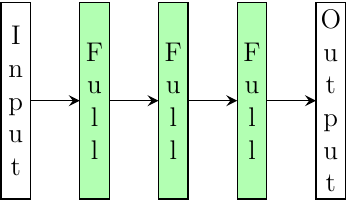}};
\end{tikzpicture}
\end{center}
\caption{\label{fig:FNN_network_3_layer} FNN with three hidden layers}
\end{figure}

and a three hidden layer FNN depicted in Figure \ref{fig:FNN_network_3_layer}

In practice one could consider a lot of other FNN architectures: Resnet\cite{He2015}, highway Nets\cite{citeulike:14070430}, DenseNets\cite{HuangGLZLW}... This could be the object of future studies.

\section{Convolutional Neural Networks} \label{sec:CNN}

Convolutional Neural Network (CNN) is a kind of network architecture particularly adapted to image classification, be it numbers or animal/car/... category. In this Section we review the novelty involved when dealing with CNN when compared to FNN as introduced in Section \ref{sec:FNN}, and do so for our regression task at hand. The most fundamental novelties are the two building blocks of CNN: convolution and pooling operations. Before presenting them, let us introduce some more notations specific to the CNN architectures

\subsection{CNN new specific notations and definitions}

In addition or in replacement to the notations introduced in Section \ref{sec:FNNdef}, we denote in the following
\begin{itemize}
\item[$\bullet$] $F_\nu$, the number of feature maps in the $\nu$'th layer.
\item[$\bullet$] $T_\nu$ and $N_\nu$, respectively the width and the height of the $\nu$'th feature map.
\item[$\bullet$] $h_{flm}^{(t)(\nu)}$ where  $\nu\in\llbracket0,N\rrbracket$ $f\in\llbracket0,F_\nu-1\rrbracket$, $l\in\llbracket0,N_\nu-1\rrbracket$ and  $m\in\llbracket0,T_\nu-1\rrbracket$,  the $\nu$'th layer components. 
\item[$\bullet$] $X^{(t)}_{flm}=h_{flm}^{(t)(0)}$ where  $f\in\llbracket0,F_0-1\rrbracket$, $l\in\llbracket0,N_0-1\rrbracket$ and  $m\in\llbracket0,T_0-1\rrbracket$, the input variables. 
\item[$\bullet$] $y^{(t)}_{flm}$ where   $f\in\llbracket0,F_N-1\rrbracket$, $l\in\llbracket0,N_N-1\rrbracket$ and  $m\in\llbracket0,T_N-1\rrbracket$, the output variables (to be predicted). 
\item[$\bullet$]  $\hat{y}^{(t)}_{flm}=h_{flm}^{(t)(N)}$ where   $f\in\llbracket0,F_N-1\rrbracket$, $l\in\llbracket0,N_N-1\rrbracket$ and  $m\in\llbracket0,T_N-1\rrbracket$, the output of the CNN. 
\item[$\bullet$] $\Theta_{f'lm}^{(\nu)f}$ for $\nu\in[0,N-1]$, $f\in [0,F_{\nu+1}-1]$, $f'\in [0,F_{\nu}-1]$, $l\in\llbracket0,N_\nu-1\rrbracket$ and $m\in\llbracket0,T_\nu-1\rrbracket$, the weight matrices.
\item[$\bullet$] $R_C$ and $S_C$, respectively, the receptive field and the stride of the convolution operation. Unless stated otherwise, these are kept the same for all the convolutions.
\item[$\bullet$] $R_P$ and $S_P$, respectively the receptive field and the stride of the pooling operation. Unless stated otherwise, these are kept the same for all the poolings.
\end{itemize}

\subsection{CNN architecture}

The CNN architecture involves convolutions (see Section \ref{sec:CNNconv}), pooling (see Section \ref{sec:CNNpool}) as well as an input, an output and fully connected (similar to those a FNN, see Section \ref{sec:FNNfc}) layers. Here is a possible CNN architecture : an input is convolved with a first weight matrix then pooled, convolved with a second weight matrix then pooled, then a fully connected operation occurs before the output is computed.

\begin{figure}[H]
\begin{center}
\begin{tikzpicture}
\node[] at (0,0) {\includegraphics[scale=0.55]{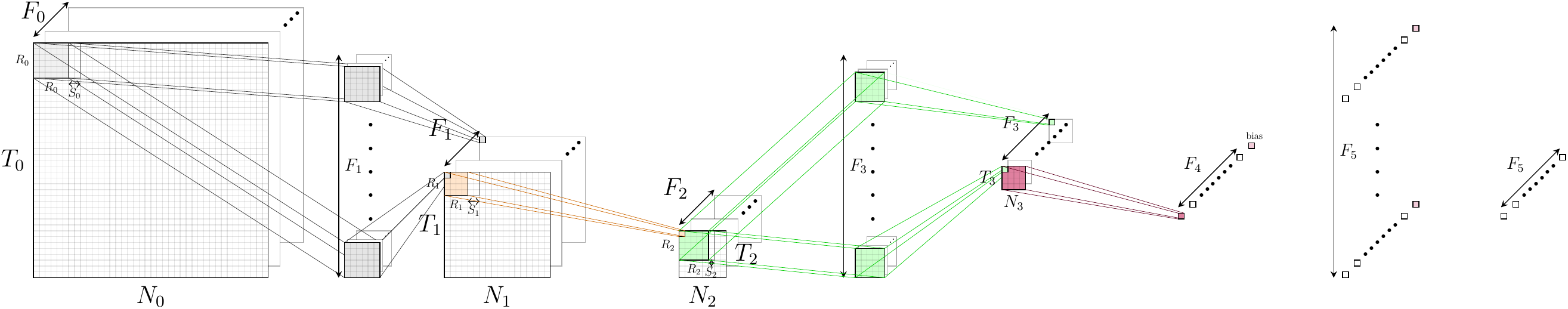}};
\end{tikzpicture}
\caption{\label{fig:lenet-CNN}A typical CNN architecture (in this case LeNet\cite{Lecun98gradient-basedlearning} inspired): convolution operations are followed by pooling operations, until the size of each feature map is reduced to one. Fully connected layers can then be introduced.}
\end{center}
\end{figure}

The fully connected layers, the output layers and the loss function are unchanged (see Sections \ref{sec:FNNfc}, \ref{sec:FNNoutput} and \ref{sec:FNNloss}). The batch normalization procedure is also used, and can easily be adapted from what has been presented in Section \ref{sec:FNNBN}. Let us now see the new and modified layers.

\subsection{Input layers} \label{sec:CNNinput}

As explained in the core of the paper, we only consider the full input for CNN. But for all purposes here, we deal with an $h^{(t)(0)}_{flm}=X^{(t)}_{f}$ input layer,  with  $f\in\llbracket0,F_0-1\rrbracket$, $l\in\llbracket0,N_0-1\rrbracket$ and  $m\in\llbracket0,T_0-1\rrbracket$.

\subsection{Convolutional layers} \label{sec:CNNconv}

The convolution operation that gives its name to the CNN is the fundamental building block of this type of network. It amounts to convolute a feature map of a hidden layer input with a weight matrix to give rise to an output feature map. The weight is really a four dimensional tensor, one dimension ($F_\nu$) being the number of feature maps of the convolutional input layer, another ($F_{\nu+1}$) the number of feature maps of the convolutional output layer. The two others give the width and the height of the receptive field. The receptive field allows one to convolute a subset instead of the whole input image. It aims at searching for similar patterns in the input image, no matter where the pattern is (translational invariance). We saw in the core of our paper the problem that this induces. The width and the height of the output image are determined by the receptive field as well as by the stride: it is simply the number of pixels by which one slides in the vertical and/or the horizontal direction before applying the convolution operation again. The central convolution formula for the $\nu$'th CNN layer (involving the $o$'th weight matrix for the $o$'th convolution operation) is
\begin{align}
a_{f\,l\,m}^{(t)(\nu)}&=\sum^{F_\nu-1}_{f'=0}\sum^{R_C-1}_{j=0}\sum^{R_C-1}_{k=0}
\Theta^{(o)f}_{f'\,j\,k}h^{(t)(\nu)}_{f'\,S_Cl+j\,S_Cm+k}\notag\\
&=\left(\Theta^{(o)T}\star h^{(t)(\nu)}\right)_{flm}\;.
\end{align}
Here $S_Cl+j$ belongs to $\llbracket0,N_\nu-1\rrbracket$ and $S_Cm+k$ to $\llbracket0,T_\nu-1\rrbracket$. This implies the following relation between $N_{\nu+1}$, $N_\nu$ and $T_{\nu+1}$, $T_\nu$ 
\begin{align}
N_{\nu+1}&=\frac{N_{\nu}-R_C}{S_C}+1\;,&
T_{\nu+1}&=\frac{T_{\nu}-R_C}{S_C}+1\;.
\end{align}
One then computes the hidden units via the ReLU activation function $g$ introduced in Section \ref{sec:FNNReLU}
\begin{align}
h_{f\,l\,m}^{(t)(\nu+1)}&=g\left(a_{f\,l\,m}^{(t)(\nu)}\right)\;.
\end{align} 

\subsection{Pooling layers}\label{sec:CNNpool}

The pooling operation, less and less used in the current state of the art CNN\cite{He2015}, is fundamentally a dimension reduction step. It amounts to take the maximum of a sub-image (characterized by the pooling receptive field $R_P$ and a stride $S_P$) of the input feature map $F_{\nu}$, to obtain an output feature map $F_{\nu+1}=F_\nu$ of width and height
\begin{align}
N_{\nu+1}&=\frac{N_{\nu}-R_P}{S_P}+1<N_\nu\;,&
T_{\nu+1}&=\frac{T_{\nu}-R_P}{S_P}+1<T_\nu\;.
\end{align}
The max pooling procedure (That we use here instead of the other possible choice, the average pooling) reads for any $\nu$
\begin{align}
a_{f\,l\,m}^{(t)(\nu)}&=\max_{j,k=\llbracket 0,R_P-1\rrbracket} h_{f\,S_P l+j\,S_Pm+k}^{(t)(\nu)}\;.
\end{align}
The hidden unit is then just
\begin{align}
h_{f\,l\,m}^{(t)(\nu+1)}&=a_{f\,l\,m}^{(t)(\nu)}\;.
\end{align}

\subsection{Architecture considered in practice}

Schematically denoting a hidden Convolution unit as
\begin{figure}[H]
\begin{center}
\begin{tikzpicture}
\node at (0,0) {\includegraphics[scale=1.8]{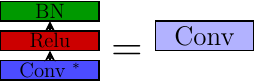}};
\end{tikzpicture}
\end{center}
\caption{\label{fig:Conv_equiv} The structure of a convolution layer}
\end{figure}

we consider in the remaining of this paper the following CNN architecture: the so-called VGG CNN\cite{DBLP:journals/corr/SimonyanZ14a}, which was the 2014 state of the art convolutional network 

\begin{figure}[H]
\begin{center}
\begin{tikzpicture}
\node at (0,0) {\includegraphics[scale=1]{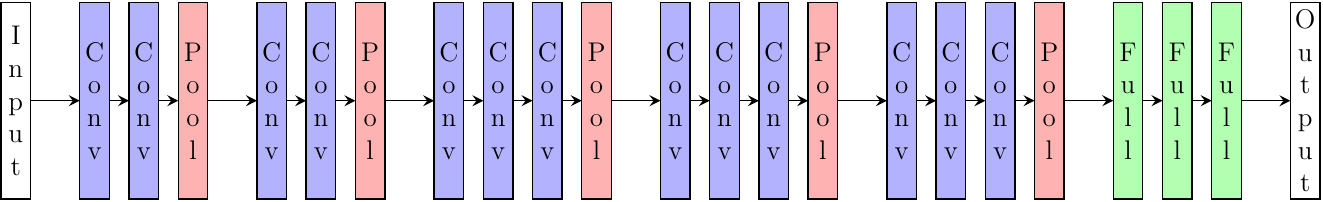}};
\end{tikzpicture}
\end{center}
\caption{\label{fig:VGG} The structure of the VGG CNN.}
\end{figure}

All the other details of the model are in the core of the paper.

\section{Recurrent Neural Networks - Long Short Term Memory} \label{sec:RNN}

In this Section, we review the third kind of Neural Network architecture used in this paper: Recurrent Neural Networks. The specificity of this kind of network is that the time dependency of the data is built into the model. We briefly present the first Recurrent Neural Network (RNN) architecture, as well as the current most popular one: the Long Short Term Memory (LSTM) Neural Network. We use the latter in this study.

\subsection{RNN new specific notations and definitions}

In contrast to the previously discussed neural networks, where we defined
\begin{align}
a^{(t)(\nu)}_{f}&= \text{ Weight Averaging } \left(h^{(t)(\nu)}_{f}\right)\;,\notag\\
h^{(t)(\nu+1)}_{f}&= \text{ Activation function } \left(a^{(t)(\nu)}_{f}\right)\;,
\end{align}
we now have hidden layers that are indexed by both a "spatial" and a "temporal" index, and the general philosophy of the RNN is (now the $a$ is usually characterized by a $c$ for cell state, this denotation, trivial for the basic RNN architecture will make more sense when we talk about LSTM networks)
\begin{align}
c^{(t)(\nu \tau )}_{f}&= \text{ Weight Averaging } \left(h^{(t)(\nu \tau-1)}_{f},h^{(t)(\nu-1\tau)}_{f}\right)\;,\notag\\
h^{(t)(\nu\tau)}_{f}&= \text{ Activation function } \left(c^{(t)(\nu \tau)}_{f}\right)\;.
\end{align}
Here, the notations are
\begin{itemize}
\item[$\bullet$] $N$, the number of layers (not counting the input) in the spatial direction.
\item[$\bullet$] $T$, the number of layers (not counting the first one) in the temporal direction (so that $T=0$ in a RNN corresponds to a standard FNN).
\item[$\bullet$] $h_{f}^{(t)(\nu\tau)}$, where  $\nu\in\llbracket0,N\rrbracket$, $\tau\in\llbracket0,T\rrbracket$ and  $f\in\llbracket0,F_\nu-1\rrbracket$,  the $\nu\tau$'th layer components. 
\item[$\bullet$] $X^{(t)(\tau)}_{f}=h_{f}^{(t)(0\tau)}$, where  $f\in\llbracket0,F_0-1\rrbracket$ and  $\tau\in\llbracket0,T\rrbracket$, the input variables (more on that in Section \ref{sec:RNNInput}). 
\item[$\bullet$] $y^{(t)(\tau)}_{f}$, where  $f\in\llbracket0,F_N-1\rrbracket$, the output variables (to be predicted). 
\item[$\bullet$]  $\hat{y}^{(t)(\tau)}_{f}$, where $f\in\llbracket0,F_N-1\rrbracket$, the output of the RNN. 
\item[$\bullet$] $\Theta_{f'}^{\nu(\nu)f}$ for $(\nu)\in[0,N-1]$, $f\in [0,F_{\nu}-1]$, $f'\in [0,F_{\nu+1}-1]$, the weight matrices in the "spatial" direction of the RNN.
\item[$\bullet$] $\Theta_{f'}^{\tau(\nu)f}$ for $(\nu)\in[1,N-1]$, $f\in [0,F_{\nu}-1]$, $f'\in [0,F_{\nu+1}-1]$, the weight matrices in the "temporal" direction of the RNN.
\end{itemize}

\subsection{RNN architecture}

\begin{figure}[H]
\begin{center}
\begin{tikzpicture}
\node at (0,0) {\includegraphics[scale=1]{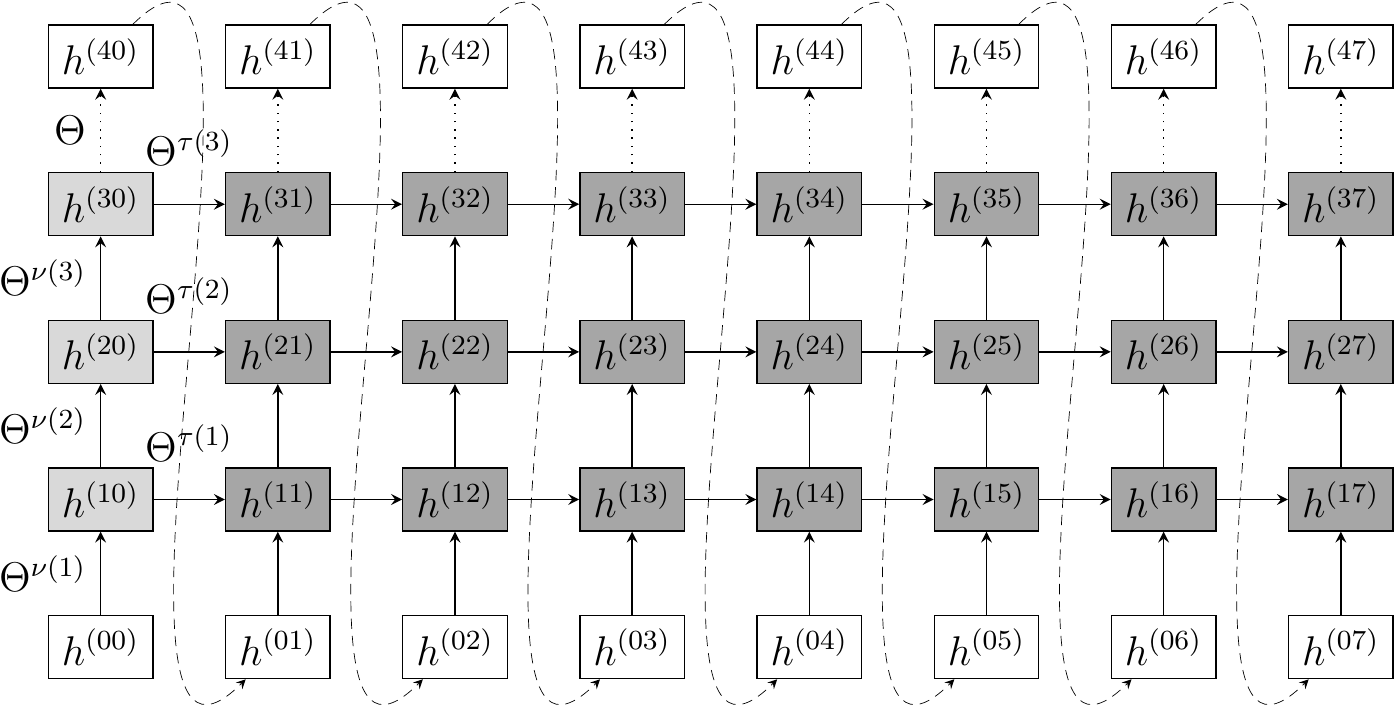}};
\end{tikzpicture}
\caption{\label{fig:RNN architecture}RNN architecture, with data propagating both in "space" and in "time". In our example, the temporal dimension is of size $T=7$ while the spatial one is of size $N=4$.}
\end{center}
\end{figure}

An example of a RNN architecture with $N=4$ spatial and $T=7$ temporal layers is depicted in Figure \ref{fig:RNN architecture}

Note that the weight matrices do not vary along the temporal direction, and this RNN only has $2(N-1)+1=7$ weight matrices (indicated in Figure \ref{fig:RNN architecture}). 

\subsection{RNN hidden layer}

The FNN formula of Section \ref{sec:FNNfc} is replaced in a RNN by (note also the change in the activation function, a standard choice in the RNN litterature\cite{GravesA2016})

\begin{align}
h^{(t)(\nu\tau)}_f&=\tanh\left(\sum_{f'=0}^{F_{{\nu-1}}-1}\Theta^{\nu(\nu)f}_{f'}
h^{(t)(\nu-1\tau)}_{f'}+\sum_{f'=0}^{F_{{\nu}}-1}\Theta^{\tau(\nu)f}_{f'}
h^{(t)(\nu\tau-1)}_{f'}\right)\;.
\end{align}

As this network has fallen out of favor because of vanishing gradient issues in backpropagation[XX], we now turn our attention to the LSTM RNN architecture that we use in the core of the paper.

\subsection{LSTM hidden layer}

\begin{figure}[H]
\begin{center}
\begin{tikzpicture}
\node[] at (0,0) {\includegraphics[scale=1.5]{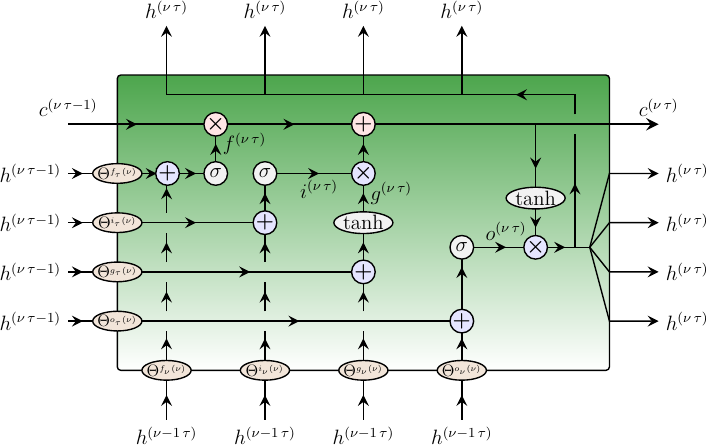}};
\end{tikzpicture}
\caption{\label{fig:Lstm1}LSTM hidden unit details}
\end{center}
\end{figure}

In a Long Short Term Memory Neural Network\cite{Gers:2000:LFC:1121912.1121915}, the state of a given unit is not directly determined by its immediate spatial and temporal neighbours. Instead, a cell state is updated for each hidden unit, and the output of this unit is a probe of the cell state. Several gates are introduced in the process: the input gate $i^{(t)(\nu\tau)}_f$ determines if we allow new information $g^{(t)(\nu\tau)}_f$ to enter into the cell state. The  output gate $o^{(t)(\nu\tau)}_f$ determines if we set or not the output hidden value to $0$, or really probe the current cell state. Finally, the forget state $f^{(t)(\nu\tau)}_f$ determines if we forget or not the past cell state. All these concepts are illustrated in Figure \ref{fig:Lstm1}

Considering all the $\tau-1$ variable values to be $0$ when $\tau=0$, we get the following formula for the input, forget and output gates:

\begin{align}
(i,f,o)^{(t)(\nu\tau)}_f&=\sigma\left(\sum_{f'=0}^{F_{{\nu-1}}-1}\Theta^{(i,f,o)_{_\nu}(\nu)f}_{f'}
h^{(t)(\nu-1\tau)}_{f'}+\sum_{f'=0}^{F{_{\nu}}-1}\Theta^{i_{_\tau}(\nu)f}_{f'}
h^{(t)(\nu\tau-1)}_{f'}\right)\;.
\end{align}
The use of the sigmoid function is the reason why the functions $i,f,o$  are called gates. Indeed, they are valued in $]0,1[$. Therefore, it allows or forbids information to pass through the next step. The cell state update is then performed in the following way
\begin{align}
g^{(t)(\nu\tau)}_f&=\tanh\left(\sum_{f'=0}^{F_{{\nu-1}}-1}\Theta^{g_{_\nu}(\nu)f}_{f'}
h^{(t)(\nu-1\tau)}_{f'}+\sum_{f'=0}^{F_{{\nu}}-1}\Theta^{g_{_\tau}(\nu)f}_{f'}
h^{(t)(\nu\tau-1)}_{f'}\right)\;,
\end{align}
and
\begin{align}
c^{(t)(\nu\tau)}_{f}&=
f^{(t)(\nu\tau)}_{f}c^{(t)(\nu\tau-1)}_{f}+i^{(t)(\nu\tau)}_{f}g^{(t)(\nu\tau)}_{f}\;.
\end{align}
So that, as announced previously, hidden state update is just a probe of the current cell state
\begin{align}
h^{(t)(\nu\tau)}_{f}&=o^{(t)(\nu\tau)}_{f}\tanh \left(c^{(t)(\nu\tau)}_{f}\right)\;.
\end{align}

There is a whole zoology of LSTM variations\cite{citeulike:14069459}, amounting to also include the cell state in the input (this is called peephole connections), forget and output update, removing some of the gates... In this paper we choose to stick with the standard LSTM formulation.

\subsection{RNN-LSTM Input Layer} \label{sec:RNNInput}

As explained in the core of the paper, we consider a full and a reduced input. But for all purposes here, we deal with an $h^{(t)(0\tau)}_{f}=X^{(t)(\tau)}_{f}$ input layer, with $f\in\llbracket0,F_0-1\rrbracket$ and $t \in \llbracket0,T_{{\rm mb}}-1\rrbracket$ and $\tau\in \llbracket0,T\rrbracket$.

\subsection{RNN-LSTM Output Layer} \label{sec:RNNoutput}

The output layer of a RNN-LSTM is similar in spirit to the FNN one and reads
\begin{align}
h^{(t)(N\tau)}_{f}&=o\left(\sum_{f'=0}^{F_{N-1}-1}\Theta^f_{f'} h^{(t)(N-1\tau)}_{f}\right)\;,
\end{align}
where the output function $o$ is the one described in Section \ref{sec:FNNoutput}.

\subsection{Change to the Loss function} \label{sec:RNNOutput}

Since the structure of the data has changed compared to a FNN/CNN, here is how the loss function reads for a regression task in a RNN-LSTM framework 
\begin{align}
J(\Theta)&=\frac{1}{2T_{{\rm mb}}}\sum_{t=0}^{T_{{\rm mb}}-1}\sum_{\tau=0}^{T-1}\sum_{f=0}^{F_N-1}
\left(h^{(t)( N\tau)}_f-y^{(t)(\tau)}_f\right)^2\;.
\end{align}

\subsection{Architecture considered in practice}

In the core of the paper, we consider a $N=2$ (hence one hidden layer) RNN-LSTM architecture, adopting the $T=20$ size to the temporal size $\mathcal{H}_{rf}$ of the full/reduced output (see the core of the paper).

\section*{Acknowledgements}

We acknowledge support from Mediamobile. We thank Philippe \textsc{Goudal} for sparking our interest in the topic. We thank Gaspard \textsc{Monge} program and EDF'lab Paris Saclay for Research initiative in industrial data science's funding.

%

\bibliographystyle{abbrv}
\bibliography{DEEP_LEARNING.bib}

\begin{thebibliography}{10}

\bibitem{agostinelli2014learning}
F.~Agostinelli, M.~Hoffman, P.~Sadowski, and P.~Baldi.
\newblock Learning activation functions to improve deep neural networks.
\newblock {\em arXiv preprint arXiv:1412.6830}, 2014.

\bibitem{caruana2001overfitting}
R.~Caruana, S.~Lawrence, and C.~L. Giles.
\newblock Overfitting in neural nets: Backpropagation, conjugate gradient, and
  early stopping.
\newblock In {\em Advances in neural information processing systems}, pages
  402--408, 2001.

\bibitem{pmlr-v40-Choromanska15}
A.~Choromanska, Y.~LeCun, and G.~B. Arous.
\newblock Open problem: The landscape of the loss surfaces of multilayer
  networks.
\newblock In P.~GrÃ¼nwald, E.~Hazan, and S.~Kale, editors, {\em Proceedings of
  The 28th Conference on Learning Theory}, volume~40 of {\em Proceedings of
  Machine Learning Research}, pages 1756--1760, Paris, France, 03--06 Jul 2015.
  PMLR.

\bibitem{Clevert2015FastAA}
D.-A. Clevert, T.~Unterthiner, and S.~Hochreiter.
\newblock Fast and accurate deep network learning by exponential linear units
  (elus).
\newblock {\em CoRR}, abs/1511.07289, 2015.

\bibitem{Deng:2016:LSM:2939672.2939860}
D.~Deng, C.~Shahabi, U.~Demiryurek, L.~Zhu, R.~Yu, and Y.~Liu.
\newblock Latent space model for road networks to predict time-varying traffic.
\newblock In {\em Proceedings of the 22Nd ACM SIGKDD International Conference
  on Knowledge Discovery and Data Mining}, KDD '16, pages 1525--1534, New York,
  NY, USA, 2016. ACM.

\bibitem{Epelbaum2017}
T.~Epelbaum.
\newblock {\em Deep Learning: Technical Introduction}.
\newblock arXiv, 2017.

\bibitem{Fazio2014}
J.~Fazio, B.~N. Wiesner, and M.~D. Deardoff.
\newblock Estimation of free-flow speed.
\newblock {\em KSCE Journal of Civil Engineering}, 18(2):646--650, Mar 2014.

\bibitem{fleischmann2004dynamic}
B.~Fleischmann, S.~Gnutzmann, and E.~Sandvo{\ss}.
\newblock Dynamic vehicle routing based on online traffic information.
\newblock {\em Transportation science}, 38(4):420--433, 2004.

\bibitem{Fouladgar2017ScalableDT}
M.~Fouladgar, M.~Parchami, R.~Elmasri, and A.~Ghaderi.
\newblock Scalable deep traffic flow neural networks for urban traffic
  congestion prediction.
\newblock {\em 2017 International Joint Conference on Neural Networks (IJCNN)},
  pages 2251--2258, 2017.

\bibitem{Gers:2000:LFC:1121912.1121915}
F.~A. Gers, J.~A. Schmidhuber, and F.~A. Cummins.
\newblock Learning to forget: Continual prediction with lstm.
\newblock {\em Neural Comput.}, 12(10):2451--2471, Oct. 2000.

\bibitem{pmlr-v9-glorot10a}
X.~Glorot and Y.~Bengio.
\newblock Understanding the difficulty of training deep feedforward neural
  networks.
\newblock In Y.~W. Teh and M.~Titterington, editors, {\em Proceedings of the
  Thirteenth International Conference on Artificial Intelligence and
  Statistics}, volume~9 of {\em Proceedings of Machine Learning Research},
  pages 249--256, Chia Laguna Resort, Sardinia, Italy, 13--15 May 2010. PMLR.

\bibitem{GravesA2016}
A.~Graves.
\newblock {\em Supervised Sequence Labelling with Recurrent Neural Networks}.
\newblock 2011.

\bibitem{citeulike:14069459}
K.~Greff, R.~K. Srivastava, J.~Koutn{\i}k, B.~R. Steunebrink, and
  J.~Schmidhuber.
\newblock {LSTM}: A search space odyssey.

\bibitem{gupta2015deep}
S.~Gupta, A.~Agrawal, K.~Gopalakrishnan, and P.~Narayanan.
\newblock Deep learning with limited numerical precision.
\newblock In {\em Proceedings of the 32nd International Conference on Machine
  Learning (ICML-15)}, pages 1737--1746, 2015.

\bibitem{Hahnloser2000}
R.~Hahnloser, R.~Sarpeshkar, M.~Mahowald, R.~J. Douglas, and S.~Seung.
\newblock Digital selection and analog amplification co-exist in an electronic
  circuit inspired by neocortex.
\newblock {\em Nature}, 405:947--951,, 06 2000.

\bibitem{He2015}
K.~He, X.~Zhang, S.~Ren, and J.~Sun.
\newblock Deep residual learning for image recognition.
\newblock 7, 12 2015.

\bibitem{HuangGLZLW}
G.~Huang, Z.~Liu, L.~Van De~Maaten, and K.~Weinberger.
\newblock Densely connected convolutional networks.
\newblock 07 2017.

\bibitem{Ioffe2015}
S.~Ioffe and C.~Szegedy.
\newblock Batch normalization: Accelerating deep network training by reducing
  internal covariate shift.
\newblock 02 2015.

\bibitem{johnson2013accelerating}
R.~Johnson and T.~Zhang.
\newblock Accelerating stochastic gradient descent using predictive variance
  reduction.
\newblock In {\em Advances in neural information processing systems}, pages
  315--323, 2013.

\bibitem{Kingma2014}
D.~Kingma and J.~Ba.
\newblock Adam: A method for stochastic optimization.
\newblock 12 2014.

\bibitem{lecun1995convolutional}
Y.~LeCun, Y.~Bengio, et~al.
\newblock Convolutional networks for images, speech, and time series.
\newblock {\em The handbook of brain theory and neural networks},
  3361(10):1995, 1995.

\bibitem{lecun2015deep}
Y.~LeCun, Y.~Bengio, and G.~Hinton.
\newblock Deep learning.
\newblock {\em Nature}, 521(7553):436--444, 2015.

\bibitem{Lecun98gradient-basedlearning}
Y.~Lecun, L.~Bottou, Y.~Bengio, and P.~Haffner.
\newblock Gradient-based learning applied to document recognition.
\newblock In {\em Proceedings of the IEEE}, pages 2278--2324, 1998.

\bibitem{LeCun:1998:EB:645754.668382}
Y.~LeCun, L.~Bottou, G.~B. Orr, and K.-R. M\"{u}ller.
\newblock Effiicient backprop.
\newblock In {\em Neural Networks: Tricks of the Trade, This Book is an
  Outgrowth of a 1996 NIPS Workshop}, pages 9--50, London, UK, UK, 1998.
  Springer-Verlag.

\bibitem{6482260}
M.~Lippi, M.~Bertini, and P.~Frasconi.
\newblock Short-term traffic flow forecasting: An experimental comparison of
  time-series analysis and supervised learning.
\newblock {\em IEEE Transactions on Intelligent Transportation Systems},
  14(2):871--882, June 2013.

\bibitem{CJS:CJS5550340307}
J.-M. Loubes, É.~Maza, M.~Lavielle, and L.~Rodriguez.
\newblock Road trafficking description and short term travel time forecasting,
  with a classification method.
\newblock {\em Canadian Journal of Statistics}, 34(3):475--491, 2006.

\bibitem{mikolov2010recurrent}
T.~Mikolov, M.~Karafi{\'a}t, L.~Burget, J.~Cernock{\`y}, and S.~Khudanpur.
\newblock Recurrent neural network based language model.
\newblock In {\em Interspeech}, volume~2, page~3, 2010.

\bibitem{MiwaTYM2004}
T.~Miwa, Y.~Tawada, T.~Yamamoto, and T.~Morikawa.
\newblock En-route updating methodology of travel time prediction using
  accumulated probe-car data.
\newblock Proc. of the 11th ITS World Congress, 2004.

\bibitem{Rosenblatt58theperceptron:}
F.~Rosenblatt.
\newblock The perceptron: A probabilistic model for information storage and
  organization in the brain.
\newblock {\em Psychological Review}, pages 65--386, 1958.

\bibitem{Sagun2014}
L.~Sagun, V.~Ugur~Guney, and Y.~Lecun.
\newblock Explorations on high dimensional landscapes.
\newblock 12 2014.

\bibitem{DBLP:journals/corr/SimonyanZ14a}
K.~Simonyan and A.~Zisserman.
\newblock Very deep convolutional networks for large-scale image recognition.
\newblock {\em CoRR}, abs/1409.1556, 2014.

\bibitem{SocherEtAl2013:CVG}
R.~Socher, J.~Bauer, C.~D. Manning, and A.~Y. Ng.
\newblock {Parsing With Compositional Vector Grammars}.
\newblock In {\em {ACL}}. 2013.

\bibitem{Srivastava:2014:DSW:2627435.2670313}
N.~Srivastava, G.~Hinton, A.~Krizhevsky, I.~Sutskever, and R.~Salakhutdinov.
\newblock Dropout: A simple way to prevent neural networks from overfitting.
\newblock {\em J. Mach. Learn. Res.}, 15(1):1929--1958, Jan. 2014.

\bibitem{citeulike:14070430}
R.~K. Srivastava, K.~Greff, and J.~Schmidhuber.
\newblock {Highway Networks}.

\bibitem{SunHongyu}
H.~Sun, H.~X. Liu, H.~Xiao, and B.~Ran.
\newblock Short term traffic forecasting using the local linear regression
  model.

\bibitem{iet:/content/conferences/10.1049/cp_20000103}
S.~Turksma.
\newblock The various uses of floating car data.
\newblock {\em IET Conference Proceedings}, pages 51--55(4), January 2000.

\bibitem{MaDaiHe:2017}
M.~Xiaolei, D.~Zhuang, H.~Zhengbing, M.~Jihui, W.~Yong, and W.~Yunpeng.
\newblock Learning traffic as images: A deep convolutional neural network for
  large-scale transportation network speed prediction.
\newblock 2017.

\bibitem{xingjian2015convolutional}
S.~Xingjian, Z.~Chen, H.~Wang, D.-Y. Yeung, W.-K. Wong, and W.-c. Woo.
\newblock Convolutional lstm network: A machine learning approach for
  precipitation nowcasting.
\newblock In {\em Advances in neural information processing systems}, pages
  802--810, 2015.

\bibitem{Yuanchang2010}
X.~Yuanchang, Z.~Kaiguang, S.~Ying, and C.~Dawei.
\newblock Gaussian processes for short-term traffic volume forecasting.
\newblock {\em Transportation Research Board of the National Academies},
  2165(-1):69--78, 2010.

\bibitem{zeiler2014visualizing}
M.~D. Zeiler and R.~Fergus.
\newblock Visualizing and understanding convolutional networks.
\newblock In {\em European conference on computer vision}, pages 818--833.
  Springer, 2014.

\bibitem{zou2005regularization}
H.~Zou and T.~Hastie.
\newblock Regularization and variable selection via the elastic net.
\newblock {\em Journal of the Royal Statistical Society: Series B (Statistical
  Methodology)}, 67(2):301--320, 2005.

\end{thebibliography}

\end{document}